\documentclass[pre,aps,10pt,superscriptaddress,twocolumn,floatfix]{revtex4-1}

\usepackage[nice]{nicefrac}
\usepackage{graphicx,color}
\usepackage{amsmath,amssymb,bm,bbm}
\usepackage{amsmath}
\usepackage{braket}
\usepackage{float}
\usepackage{placeins}

\usepackage{physics} 
\usepackage[normalem]{ulem} 
\usepackage[T1]{fontenc}
\usepackage[utf8]{inputenc}

\newcommand{\oo}{o}

\usepackage[plainpages=false,pdfpagelabels,colorlinks=true,linkcolor=red,urlcolor=blue,citecolor=blue,pdftitle={},pdfauthor={},pdfdisplaydoctitle=true,pdfduplex=DuplexFlipLongEdge]{hyperref}

\definecolor{darkred}{rgb}{0.90,0.2,0.2}
\definecolor{darkgreen}{rgb}{0,0.60,.2}
\definecolor{darkblue}{rgb}{0.1,0.3,1}
\definecolor{grey}{cmyk}{0,0,0,0.25}
\definecolor{orange}{cmyk}{0,0.6,0.8,0}

\begin{document}

\title{Normal weak eigenstate thermalization}

\author{Patrycja  \L yd\.{z}ba}
\affiliation{Institute of Theoretical Physics, Wroclaw University of Science and Technology, 50-370 Wroc{\l}aw, Poland}
\author{Rafał Świętek}
\affiliation{Department of Theoretical Physics, J. Stefan Institute, SI-1000 Ljubljana, Slovenia}
\affiliation{Department of Physics, Faculty of Mathematics and Physics, University of Ljubljana, SI-1000 Ljubljana, Slovenia\looseness=-1}
\author{Marcin Mierzejewski}
\affiliation{Institute of Theoretical Physics, Wroclaw University of Science and Technology, 50-370 Wroc{\l}aw, Poland}
\author{Marcos Rigol}
\affiliation{Department of Physics, The Pennsylvania State University, University Park, Pennsylvania 16802, USA}
\author{Lev Vidmar}
\affiliation{Department of Theoretical Physics, J. Stefan Institute, SI-1000 Ljubljana, Slovenia}
\affiliation{Department of Physics, Faculty of Mathematics and Physics, University of Ljubljana, SI-1000 Ljubljana, Slovenia\looseness=-1}

\begin{abstract}
Eigenstate thermalization has been numerically shown to occur for few-body observables in a wide range of nonintegrable interacting models. For intensive sums of few-body observables, a weaker version of eigenstate thermalization known as weak eigenstate thermalization has been proved to occur in general. Here, we unveil a novel weak eigenstate thermalization phenomenon that occurs in quadratic models exhibiting quantum chaos in the single-particle sector (quantum-chaotic quadratic models) and in integrable interacting models. In such models, we argue that few-body observables that have a properly defined system-size independent norm are guaranteed to exhibit at least a polynomially vanishing variance (over the entire many-body energy spectrum) of the diagonal matrix elements, a phenomenon we dub {\it normal} weak eigenstate thermalization. We prove that normal weak eigenstate thermalization is a consequence of single-particle eigenstate thermalization, i.e., it can be viewed as a manifestation of quantum chaos at the single-particle level. We report numerical evidence of normal weak eigenstate thermalization for quantum-chaotic quadratic models such as the 3D Anderson model in the delocalized regime and the power-law random banded matrix model, as well as for the integrable interacting spin-$\frac{1}{2}$ XYZ and XXZ models. 
\end{abstract}

\maketitle

\section{Introduction}

The eigenstate thermalization hypothesis (ETH) explains why, in general, thermalization occurs in isolated quantum systems~\cite{deutsch_91, srednicki_94, rigol_dunjko_08}. The ETH can be written as an ansatz for the matrix elements of observables (few-body operators) $\hat O$ in the energy eigenstates $\{|\Omega\rangle\}$~\cite{srednicki_99, dalessio_kafri_16}:
\begin{equation} \label{def_eth}
     \langle \Omega | \hat O | \Gamma \rangle = O(\bar E) \delta_{\Omega\Gamma} + \frac{1}{\sqrt{e^{S(\bar E)}}} f_O(\bar E, \omega) R_{\Omega\Gamma}^O \, ,
\end{equation}
where $\bar E = (E_\Omega + E_\Gamma)/2$ is the average energy of the pair of eigenstates $|\Omega\rangle$ and $|\Gamma\rangle$, $\omega=E_\Omega - E_\Gamma$ is their energy difference, $O(\bar E)$ and $f_O(\bar E,\omega)$ are smooth functions of their arguments, $S(\bar E)$ is the thermodynamic entropy at energy $\bar E$, and $R_{\Omega\Gamma}^O$ are random numbers with zero mean and unit variance. Since $S(\bar E)$ is an extensive quantity, it follows from this ansatz that the fluctuations of the diagonal matrix elements about their mean $O(\bar E)$, as well as the magnitude of the off-diagonal matrix elements, decay exponentially with increasing the system size. An isolated quantum system is said to exhibit eigenstate thermalization when the number of matrix elements of observables (in the energy eigenstates) that are not described by Eq.~(\ref{def_eth}) vanishes in the thermodynamic limit.

The ETH ansatz~\eqref{def_eth} is expected to describe the matrix elements of few-body operators in the eigenstates of nonintegrable (quantum-chaotic) interacting models~\cite{dalessio_kafri_16, mori_ikeda_18, deutsch_18}. This expectation is supported by the fact that eigenstate thermalization has been numerically found to occur in a wide range of quantum-chaotic Hamiltonians with short-range interactions~\cite{rigol_dunjko_08, rigol_09a, rigol09, santos_rigol_10b, steinigeweg_herbrych_13, khatami_pupillo_13, beugeling_moessner_14, sorg14, steinigeweg_khodja_14, kim_ikeda_14, beugeling_moessner_15, mondaini_fratus_16, mondaini_rigol_17, yoshizawa_iyoda_18, nation_porras_18, khaymovich_haque_19, jansen_stolpp_19, leblond_mallayya_19, mierzejewski_vidmar_20, brenes_leblond_20, brenes_goold_20, santos_perezbernal_20, richter_dymarsky_20, leblond_rigol_20, schoenle_jansen_21, sugimoto_hamazaki_21, noh_21, wang_lamann_22}. Extensions to describe higher-order correlations have also been discussed~\cite{foini_kurchan_19, chan_deluca_19, murthy_srednicki_19b, brenes_pappalardi_21, pappalardi_foini_22, pappalardi_fritzsch_23, wang_richter_23}. On the other hand, eigenstate thermalization does not occur in quadratic nor in integrable interacting models~\cite{rigol_dunjko_08, rigol_09a, rigol09, santos_rigol_10b, Cassidy_2011, vidmar16}, in which the fluctuations of the diagonal matrix elements about their mean decay at most polynomially with increasing system size~\cite{biroli_kollath_10, ikeda_watanabe_13, steinigeweg_herbrych_13, beugeling_moessner_14, alba_15, mori_16, leblond_mallayya_19, mierzejewski_vidmar_20, leblond_rigol_20, zhang_vidmar_22, essler_klerk_23}. Other examples of absence of eigenstate thermalization include many-body localized regimes~\cite{luitz_16, serbyn_papic_17, colmenarez_mcclarty_19, panda_scardicchio_20, luitz_khaymovich_20}, systems with long-range interactions~\cite{russomanno_fava_21, sugimoto_hamazaki_22}, quantum many-body scars~\cite{serbyn_abanin_21, moudgalya_bernevig_22, channdran_iadecola_23, moudgalya_motrunich_22}, and Hilbert space fragmentation~\cite{razkovsky_sala_20, moudgalya_motruninch_22,PhysRevB.101.125126}. When the fluctuations of the diagonal matrix elements of an observable decay polynomially, as in many quadratic and integrable interacting models~\cite{biroli_kollath_10, ikeda_watanabe_13, steinigeweg_herbrych_13, beugeling_moessner_14, alba_15, mori_16, leblond_mallayya_19, mierzejewski_vidmar_20, leblond_rigol_20, zhang_vidmar_22, essler_klerk_23}, or faster~\cite{PhysRevE.104.024135, PhysRevX.13.031013, Reimann_2018, PhysRevLett.119.030601, PhysRevLett.123.147201} with increasing system size, which means that the fraction of diagonal matrix elements that differ from the microcanonical average vanishes in the thermodynamic limit, that observable is said to exhibit {\it weak} eigenstate thermalization~\cite{biroli_kollath_10}. It is important to stress that, in systems that exhibit weak eigenstate thermalization, the number of diagonal matrix elements that differ from the microcanonical average can be exponentially large in the system size, and that this is actually the case in quadratic and integrable interacting models~\cite{Cassidy_2011, vidmar16, lydzba_mierzejewski_23}.

For quadratic and integrable interacting Hamiltonians, it is understood that because of the lack of eigenstate thermalization (specifically, because of the exponentially large number of nonthermal states), observables after quantum quenches in general do not thermalizate. Instead, their long-time averages are described by generalized Gibbs ensembles (GGEs)~\cite{rigol_dunjko_07, cazalilla_06, rigol_muramatsu_06, iucci_cazalilla_09, Cassidy_2011, calabrese_essler_11, Gramsch_2012, calabrese_essler_12b, essler_evangelisti_12, Ziraldo_2012, Ziraldo_2013, He_2013, caux_essler_13, wright_rigol_14, wouters_denardis_14, pozsgay_mestyan14, ilievski_denardis_15, vidmar16, calabrese_essler_review_16}. While the long-time averages of both quadratic and integrable interacting models are described by GGEs, there is a distinction between those models when it comes to the fluctuations of observables about their time averages. Equilibration occurs when those fluctuations become vanishingly small with increasing system size. In quadratic models, the single-particle eigenstates are often localized in a physically relevant single-particle basis, e.g., in quasimomentum space in the presence of discrete lattice translational invariance (single-particle states are Bloch states), or in position space in the presence of uncorrelated disorder (as a consequence of Anderson localization). Because of this, in many-body sectors of localized quadratic models some observables equilibrate while others fail to do so~\cite{Gramsch_2012, Ziraldo_2012, Ziraldo_2013, wright_rigol_14}, see also Refs.~\cite{cramer_dawson_08, gluza_krumnow_16, murthy19, gluza_eisert_19}. On the other hand, in integrable interacting models (which are usually clean models in one dimension) few-body observables have been found to generally equilibrate~\cite{rigol_09a, rigol09, wouters_denardis_14, pozsgay_mestyan14, ilievski_denardis_15}. The question of equilibration has also been extensively studied in the context of typicality~\cite{Reimann_2008, linden_09, Short_2011, Short_2012, Reimann_2012, PhysRevE.87.012106, Eisert_2015, Gluza_2016, PhysRevE.93.062107, Gluza_2019}.

In recent years, we have explored the properties of a class of quadratic models that to some extent mirrors the properties of integrable interacting models. Those models exhibit quantum chaos and eigenstate thermalization at the single-particle level~\cite{lydzba_zhang_21, Tokarczyk_2023}. Namely, the distribution of spacings of nearest-neighbor energy levels in the single-particle spectrum is Wigner-Dyson, and the matrix elements of observables in the single-particle energy eigenstates comply with the ETH ansatz in the single-particle sector [equivalent to Eq.~\eqref{def_eth}, see Eq.~(\ref{O_sp})]. We refer to such models as quantum-chaotic quadratic (QCQ) models. In Ref.~\cite{lydzba_mierzejewski_23}, one-body observables and their products were proved to always equilibrate in many-body sectors of QCQ models. Because there is an exponentially large number (but a vanishing fraction in the thermodynamic limit) of many-body QCQ Hamiltonian eigenstates whose diagonal matrix elements are nonthermal, observables after equilibration are described by a GGE rather than by a thermal ensemble~\cite{lydzba_mierzejewski_23}.

In this work we use QCQ models to unveil a novel form of weak eigenstate thermalization. A form that, for few-body observables (as defined in Sec.~\ref{sec:fewbody}), is only guaranteed to arise in QCQ and (per our numerical results) in integrable interacting models. Previous works proved that weak eigenstate thermalization is a phenomenon that occurs for intensive few-body observables that are sums of local operators~\cite{biroli_kollath_10, mori_16, iyoda17}. As pointed out in Ref.~\cite{biroli_kollath_10}, even in very simple quadratic models, the fluctuations of the diagonal matrix elements of such operators in the energy eigenbasis vanish (in any microcanonical energy window) at least polynomially with increasing the system size. Related to those results, we note that the vanishing of the fluctuations of the diagonal matrix elements of few-body observables that are intensive sums of local operators when calculated over the entire energy spectrum, namely, ``infinite-temperature'' weak eigenstate thermalization, is a direct consequence of the dependence of the norm of such operators on the system size~\cite{mierzejewski_vidmar_20}. On the other hand, for QCQ and integrable interacting models, here we unveil a stricter form of ``infinite-temperature'' weak eigenstate thermalization that always occurs for few-body operators with a properly defined system-size independent norm, namely, for normalized few-body operators. We dub it normal weak eigenstate thermalization as the fluctuations of the diagonal matrix elements vanish at least polynomially while the relevant norm does not.

Specifically, we prove for one-body observables, and argue for few-body observables, that normal weak eigenstate thermalization is guaranteed to occur whenever eigenstate thermalization occurs in the single-particle sector of the Hamiltonian. We provide numerical evidence of normal weak eigenstate thermalization for QCQ models such as the 3D Anderson model in the delocalized regime and the power-law random banded matrix model, as well as for the integrable interacting spin-$\frac{1}{2}$ XYZ and XXZ models. In contrast, for quadratic models that exhibit localization in either position or momentum space, we provide numerical evidence that normal weak eigenstate thermalization does not occur for classes of observables that can be identified a priori.

The presentation is organized as follows. In Sec.~\ref{sec:observables}, we introduce the few-body observables under investigation and their sums, and discuss the importance of the normalization. In Sec.~\ref{sec:quadraticH}, we introduce the QCQ and localized quadratic Hamiltonians that are studied in this work. In Sec.~\ref{sec:spet_weakETH}, we establish analytically the connection between single-particle eigenstate thermalization and normal weak eigenstate thermalization for one-body observables in QCQ models. We study numerically the scaling of the variances of the matrix elements of one-body observables in QCQ models in Sec.~\ref{sec:chaotic}, and in localized quadratic models that do not exhibit quantum chaos in their single-particle sector in Sec.~\ref{sec:localized}. The generalization to two-body and higher few-body observables is discussed in Sec.~\ref{sec:tb_weakETH}. Normal weak eigenstate thermalization in interacting integrable Hamiltonians is studied numerically in Sec.~\ref{sec:interacting}. We conclude with a summary and discussion of our results in Sec.~\ref{sec:conclusions}.

\section{Observables and normalization} \label{sec:observables}

We study one-body (two-point) and two-body (four-point) observables. General one-body observables can be written as 
\begin{equation} \label{def_oA}
    \hat o_{\rm I}=\sum_{\alpha,\beta=1}^V o^{(\rm I)}_{\alpha\beta}\, \hat c_{\alpha}^\dagger \hat c^{}_\beta \,,
\end{equation}
where $\hat c^{\dagger}_\alpha$ ($\hat c^{}_\alpha$) creates (annihilates) a spinless fermion in a single-particle state $|\alpha \rangle$, while general two-body observables can be written as
\begin{equation} \label{def_oB}
    \hat o_{\rm II}=\sum_{\alpha,\beta,\gamma,\delta=1}^V o^{(\rm II)}_{\alpha\beta\gamma\delta}\, \hat c_{\alpha}^\dagger \hat c_\beta^\dagger \hat c^{}_{\gamma} \hat c^{}_{\delta} \,.
\end{equation}
Our focus is on specific forms of those observables to which we refer as one- and two-body observables, see Sec.~\ref{sec:fewbody}, and their sums, see Sec.~\ref{sec:sums}. 

We study quadratic models of spinless fermions with $V$ sites in one-dimensional (1D) and three-dimensional (3D) lattices, where $V=L$ in 1D lattices and $V=L^3$ in 3D lattices, with $L$ being the linear number of sites. We also study integrable interacting chains, for which we refer to the number of lattice sites as $L$.

\subsection{One- and two-body observables} \label{sec:fewbody}

We call one-body observables the operators $\hat o_{\rm I}$ from Eq.~(\ref{def_oA}) that, in the thermodynamic limit ($V \to \infty$), can be expressed as \textit{finite sums} of products of one creation and one annihilation operator in some single-particle basis. Formally, our one-body observables have rank $\mathbf{O}(1)$ in the single-particle Hilbert space, where the rank is the number of nondegenerate eigenvalues. 

Specifically, we study the nearest neighbor hopping operator,
\begin{equation}
\label{eq_h1}
\hat{h}^{(1)}_i = \hat{c}_{i}^\dagger \hat{c}^{}_{i+1} + \hat{c}_{i+1}^\dagger \hat{c}^{}_{i} \,,
\end{equation}
and the next-nearest neighbor hopping operator,
\begin{equation}
\hat{h}^{(2)}_i = \hat{c}_{i}^\dagger \hat{c}^{}_{i+2} + \hat{c}_{i+2}^\dagger \hat{c}^{}_{i}\,,
\end{equation}
where we use $i$ to label the lattice sites. In 3D lattices, $i\equiv \{x,y,z\}$, $i+1\equiv \{x+1,y,z\}$, and $i+2\equiv \{x+1,y+1,z\}$. We always compute these operators at a properly chosen site $i^*$, and report results for $\hat{h}_1\equiv \hat{h}^{(1)}_{i^*}$ and $\hat{h}_2\equiv \hat{h}^{(2)}_{i^*}$. For the models with periodic boundary conditions, the chosen site $i^*$ is not important. For the models with open boundary conditions, which are 1D models in this work, we take the site $i^*$ to be at the center of the chain.

The operators $\hat{h}_1$ and $\hat{h}_2$ are one-body operators that are local in position space. We also study a one-body operator that is nonlocal in position space, the occupation of the zero quasimomentum mode:
\begin{equation}
    \hat m_0 = \hat f_0^\dagger \hat f^{}_0 \,,
\end{equation}
where $\hat f^{}_0 = \sum_{j=1}^V \hat c^{}_j/\sqrt{V}$ in a lattice with $V$ sites.

We call two-body observables the operators $\hat o_{\rm II}$ from Eq.~(\ref{def_oB}) that can be expressed as \textit{finite sums} of products of two creation and two annihilation operators in some single-particle basis. Specifically, we study the nearest neighbor density-density operator
\begin{equation}
\label{eq_o1}
\hat{o}^{(1)}_i = \hat{n}_{i} \hat{n}_{i+1} =\hat{c}_{i}^\dagger\hat{c}_{i+1}^\dagger\hat{c}^{}_{i+1}\hat{c}^{}_{i}\,,
\end{equation}
the next-nearest neighbor density-density operator,
\begin{equation}
\label{eq_o2}
\hat{o}^{(2)}_i = \hat{n}_{i} \hat{n}_{i+2} =\hat{c}_{i}^\dagger\hat{c}_{i+2}^\dagger\hat{c}^{}_{i+2}\hat{c}^{}_{i}\,,
\end{equation}
where
\begin{equation}
    \hat n_i = \hat c_i^\dagger \hat c^{}_i \,,
\end{equation} 
and the correlated (pair) hopping,
\begin{equation}
\label{eq_o3}
\hat{o}^{(3)}_i = \hat c_i^\dagger \hat c_{i+3}^\dagger \hat c^{}_{i+2} \hat c^{}_{i+1} + {\rm H.c.} \,.
\end{equation}
For the correlated hopping in 3D lattices, $i\equiv \{x,y,z\}$, $i+1\equiv \{x+1,y,z\}$, $i+2\equiv \{x+2,y,z\}$, and $i+3\equiv \{x+3,y,z\}$. Like for the one-body operators, we compute the two-body operators at a properly chosen site $i^*$, and report results for $\hat{o}_1\equiv \hat{o}^{(1)}_{i^*}$, $\hat{o}_2\equiv \hat{o}^{(2)}_{i^*}$, and $\hat{o}_3\equiv \hat{o}^{(3)}_{i^*}$. For models with periodic boundary conditions, the chosen site $i^*$ is not important. For 1D models with open boundary conditions, we take $i^*$ to be at the center of the chain.

One can similarly define three-body and higher few-body operators following this procedure. Those are the kind of operators that we have in mind whenever we mention few-body observables in this work.

\subsection{Sums of one- and two-body observables} \label{sec:sums}

We define sums of one-body (two-body) observables as the operators $\hat o_{\rm I}$ in Eq.~(\ref{def_oA}) [$\hat o_{\rm II}$ in Eq.~(\ref{def_oB})] that, in the thermodynamic limit ($V \to \infty$), cannot be expressed as finite sums of products of two (four) creation and annihilation operators in some single-particle basis. In the case of one-body observables, such operators have rank $\mathbf{O}(V)$ in the single-particle Hilbert space. 

Specifically, as sums of one-body observables, we study the sums of nearest and next-nearest neighbor hopping operators,
\begin{equation} \label{def_T1_T2}
    \hat{T}_1=\frac{1}{\sqrt{V}}\sum_{i=1}^{V} \hat{h}^{(1)}_i\,,\quad
    \hat{T}_2=\frac{1}{\sqrt{V}}\sum_{i=1}^{V} \hat{h}^{(2)}_i\,,
\end{equation}
and, as sums of two-body observables, we study the sums of nearest and next-nearest neighbor density-density operators,
\begin{equation} \label{def_O1_O2}
    \hat{O}_1=\frac{1}{\sqrt{V}}\sum_{i=1}^{V} \hat{o}^{(1)}_i\,,\quad
    \hat{O}_2=\frac{1}{\sqrt{V}}\sum_{i=1}^{V} \hat{o}^{(2)}_i\,.
\end{equation}
A common property of the sums of one- and two-body observables considered here is the prefactor $1/\sqrt{V}$, which ensures that the norm of their traceless counterparts [defined in Eq.~(\ref{def_norm_o_mb})] is ${\bf O}(1)$. We discuss the consequences of using the traditional $1/V$ normalization, which ensures the sums are intensive operators, in Sec.~\ref{sec:normalwETH}.

One can similarly define sums of three-body and higher few-body operators following this procedure. Those are the kind of operators that we have in mind whenever we mention sums of few-body observables in this work.

\subsection{Variance of diagonal matrix elements}

We are interested in how the fluctuations of the diagonal matrix elements of observables $O_{\Omega\Omega} \equiv \langle \Omega|\hat O|\Omega\rangle$ in the many-body energy eigenstates $|\Omega\rangle$ across the energy spectrum scale with the system size. Those fluctuations can be quantified by the variance
\begin{equation} \label{def:variance}
    \sigma^2 = \frac{1}{D} \sum_{\Omega=1}^D \left[O_{\Omega\Omega} - O_{\rm mic}(E_\Omega)\right]^2 \,,
\end{equation}
where $D$ is the dimension of the many-body Hilbert space, which is finite for models considered here in finite lattice sizes, and $O_{\rm mic}(E_\Omega)$ is the expectation value of the observable in the microcanonical ensemble at energy $E_\Omega$. The latter can be computed as a running average~\cite{beugeling_moessner_14, mondaini_rigol_17, jansen_stolpp_19, mierzejewski_vidmar_20} 
\begin{equation} \label{def:micave}
O_{\rm mic}(E_\Omega)=\frac{1}{\cal N} \sum_{E_{\Omega'} \in {\cal M}} O_{\Omega'\Omega'},
\end{equation} 
where ${\cal M} = \{E_{\Omega'};\,  E_\Omega - \Delta/2 \leq E_{\Omega'} \leq  E_\Omega + \Delta/2\}$ with a width $\Delta$ about $E_\Omega$, and $\cal N$ is the number of energy eigenstates in the microcanonical window. We emphasize that the variance $\sigma^2$ in Eq.~(\ref{def:variance}) is calculated over the entire many-body energy spectrum. It can be thought of as an ``infinite-temperature'' variance.

\subsection{Norm of few-body operators}

The operators $\hat O$ of interest in this work, associated to physical observables, are few-body Hermitian operators. Their Hilbert-Schmidt norm,
\begin{equation}
\label{eq_textbook}
||\hat{O}||^2_\text{HS} \equiv \Tr \left[\hat{O}^\dagger\hat{O}\right] = \Tr \left[\hat{O}^2\right]=\sum_{\Omega,\Gamma=1}^{D}O_{\Omega\Gamma}^2,
\end{equation}
is proportional to the dimension of the Hilbert space $D$, e.g., for a site occupation operator $\hat n_i$, $||\hat{n}_i||^2_\text{HS}=D/2$. Such a scaling with $D$ is implicitly assumed when writing the ETH ansatz, so that $O(\bar E)$ and $f_O(\bar E, \omega)$ can be ${\bf O}(1)$.

To focus on prefactors in the normalization of few-body operators that depend polynomially on the number of lattice sites, which are central to weak eigenstate thermalization, we remove the trivial dependence of the Hilbert-Schmidt norm of few-body operators on the dimension of the Hilbert space $D$. Namely, we define the norm of a few-body operator $\hat{O}$ to be:
\begin{equation}
\label{eq_norm2}
||\hat{O}||^2 \equiv \frac{1}{D} ||\hat{O}||^2_\text{HS} =\frac{1}{D}\sum_{\Omega,\Gamma=1}^{D}O_{\Omega\Gamma}^2.
\end{equation}
Note that $||\hat{O}||^2$ in Eq.~(\ref{eq_norm2}) satisfies all of the axioms of a norm. Moreover, the variance $\sigma^2$ in Eq.~(\ref{def:variance}) is upper bounded by $||\hat{O}||^2$. As shown in Ref.~\cite{mierzejewski_vidmar_20}:
\begin{equation}
\label{eq_bound}
     \sigma^2 \le \frac{1}{D} \sum_{\Omega=1}^D \left[O_{\Omega\Omega} - f(E_\Omega)\right]^2,
\end{equation}
where $f(E_\Omega)$ is an arbitrary smooth function of the energy $E_\Omega$. The equality is attained when $f(E_\Omega)=O_\text{mic}(E_\Omega)$. For $f(E_\Omega)=0$, we find
\begin{equation} \label{bound_sigma}
\begin{split}
    \sigma^2 & \le \frac{1}{D} \sum_{\Omega=1}^D \left[O_{\Omega\Omega}\right]^2 \le \frac{1}{D} \sum_{\Omega=1}^D \left[O_{\Omega\Omega}\right]^2 + \frac{1}{D} \sum_{\Omega\neq\Gamma}O_{\Omega\Gamma}^2\\
    & =||\hat{O}||^2,
\end{split}
\end{equation}
so that $0\le\sigma^2/||\hat{O}||^2\le 1$. Both bounds can be saturated for physically relevant operators. In particular, $\sigma^2/||\hat{O}||^2=0$ for observables in macroscopic systems that obey the ETH, while $\sigma^2/||\hat{O}||^2=1$ for conserved structureless observables in integrable models (e.g., the energy current in the XXZ chain).

Modifying an operator by adding a constant neither changes its physical meaning nor the description of its matrix elements via the ETH ansatz [see Eq.~(\ref{def_eth})]. This constant can be trivially absorbed in the function $O(\bar E)$. However, such a constant can change the norm in Eq.~(\ref{eq_norm2}). In order to eliminate potentially ambiguous constants from our considerations, we normalize the traceless counterparts of the operators considered. In other words, we first project the operator $\hat O$ onto the subspace of traceless operators, $\mathcal{P}\hat{O}=\hat{O}-(\text{Tr}[\hat{O}]/D)\hat{I}$, where $\mathcal{P}$ and $\hat{I}$ are the projection superoperator and the identity operator, respectively, associated to the many-body Hilbert space; and then compute the norm $||\mathcal{P}\hat{O}||$. For simplicity, we introduce the notation 
\begin{eqnarray} \label{def_norm_o_mb}
    ||\hat{O}||^2_\text{mb} &\equiv&||\mathcal{P}\hat{O}||^2=||\hat{O}||^2-
    \left(\frac{1}{D}\text{Tr}[\hat{O}]\right)^2 \\  &=& \frac{1}{D}\sum_{\Omega=1}^{D} \langle\Omega|\hat{O}^2|\Omega\rangle-\left(\frac{1}{D}\sum_{\Omega=1}^{D} \langle\Omega|\hat{O}|\Omega\rangle\right)^2.\nonumber
\end{eqnarray}
From now on, we refer to $||\hat{O}||_\text{mb}$ as the norm of operator $\hat{O}$ in the many-body Hilbert space. Readers should keep in mind that we mean the norm of the traceless version of $\hat O$, which is invariant under a shift by an arbitrary constant, in the many-body Hilbert space. 

Next, we note that selecting the smooth function in Eq.~(\ref{eq_bound}) to be the constant $f(E_\Omega)=\frac{1}{D}\sum_{\Gamma}O_{\Gamma\Gamma}$, we obtain
\begin{equation}
    \sigma^2\le \frac{1}{D} \sum_{\Omega=1}^D |O_{\Omega\Omega}|^2 - \left( \frac{1}{D} \sum_{\Omega =1}^D O_{\Omega\Omega} \right)^2.
\end{equation}
If we rewrite Eq.~\eqref{def_norm_o_mb} by adding an identity operator to the first term, $\hat{O}^2 = \hat{O}\hat I \hat{O}=\sum_\Gamma \hat{O} |\Gamma\rangle\langle \Gamma | \hat{O}$, we arrive at
\begin{equation}\label{eq:HSN_tmp}
    ||\hat O||^2_\text{mb}  = \frac{1}{D} \sum_{\Omega, \Gamma=1}^D |O_{\Omega\Gamma}|^2 - \left( \frac{1}{D} \sum_{\Omega =1}^D O_{\Omega\Omega} \right)^2.
\end{equation}
Hence, the variance $\sigma^2$ is also upper bounded by the introduced norm $||\hat{\mathcal{O}}||^2_\text{mb}$:
\begin{equation}\label{eq:wETHsums}
    \sigma^2 \leq ||\hat{\mathcal{O}}||^2_\text{mb} - \frac{1}{D} \sum_{\Omega\neq\Gamma} |O_{\Omega\Gamma}|^2 \le ||\hat{\mathcal{O}}||^2_\text{mb} \,.
\end{equation}

In quadratic systems, the matrix elements of few-body operators in many-body energy eigenstates $\left\{|\Omega\rangle\right\}$ can be written in terms of the matrix elements of one-body operators in single-particle energy eigenstates $\left\{|\omega\rangle\right\}$. Therefore, we also study the properties of the latter. Since the dimension of the Hilbert space of the single-particle sector is $V$, in contrast to $D=2^V$ for the many-body Hilbert space, next we explicitly rewrite the equations that we will use in the single-particle derivations replacing $D\rightarrow V$ and adding the sub-index ``${\rm sp}$''. 

The variance of the diagonal matrix elements reads,
\begin{equation}
    \sigma^2_{\rm sp} = \frac{1}{V} \sum_{\omega=1}^V \left[O_{\omega\omega} - O_{\rm mic}(\epsilon_\omega)\right]^2 \,,
\end{equation}
with $\epsilon_\omega$ being the single-particle eigenenergies and $O_{\rm mic}(\epsilon_\omega)$ being the microcanonical averages in the single-particle sector. As in the many-body case, $\sigma^2_{\rm sp}$ is bounded from above by the norm in the single-particle sector 
\begin{equation}
\label{eq_norm_sp}
    \widetilde{||\hat{O}||^{2}} \equiv \frac{1}{V} \Tr_\text{sp} \left[\hat{O}^2\right]=\frac{1}{V}\sum_{\omega,\gamma=1}^{V}O_{\omega\gamma}^2.
\end{equation}
We further define $||\hat{O}||_\text{sp}$ as the norm of the projected operator $\mathcal{P}_\text{sp}\hat{O}=\hat{O}-(\text{Tr}_\text{sp}[\hat{O}]/V)\hat{I}_\text{sp}$, where $\mathcal{P}_\text{sp}$ and $\hat{I}_\text{sp}$ are the projection superoperator and the identity operator, respectively, associated to the single-particle Hilbert space, i.e.,
\begin{eqnarray}
\label{def_O_sp}
     ||\hat{O}||^2_\text{sp}&=&\widetilde{||\mathcal{P}_\text{sp}\hat{O}||^2}=\widetilde{||\hat{O}||}^2-\left(\frac{1}{V}\text{Tr}_\text{sp}[\hat{O}]\right)^2  \\ &=& \frac{1}{V}\sum_{\omega=1}^{V}\langle\omega|\hat{O}^2|\omega\rangle-\left(\frac{1}{V}\sum_{\omega=1}^{V}\langle\omega|\hat{O}|\omega\rangle\right)^2 \,.\nonumber
\end{eqnarray}
In Eqs.~(\ref{eq_norm_sp}) and~(\ref{def_O_sp}), the trace is taken over single-particle energy eigenstates. In what follows, we call $||\hat{O}||^2_\text{sp}$ the norm of $\hat{O}$ in the single-particle sector. Readers should keep in mind that we mean the norm of its traceless counterpart $\mathcal{P}_\text{sp}\hat{O}$ in the single-particle sector.

To close this section, we note that the matrix elements of one-body operators in the many-body and single-particle Hilbert spaces are related via
\begin{equation}
\label{eq_relation}
\langle \Omega |\hat O | \Omega' \rangle=\sum_{\omega,\omega'=1}^{V} \langle \omega |\hat O | \omega' \rangle \langle \Omega |\hat f^{\dagger}_{\omega} 
\hat f^{}_{\omega'} | \Omega' \rangle.
\end{equation}
The norms $||...||_{\text{mb}}$ and $||...||_{\text{sp}}$ are defined for $D\times D$ matrices $\langle \Omega |\hat O | \Omega' \rangle$ and $V\times V$ matrices $\langle \omega |\hat O | \omega' \rangle$, respectively.

\subsection{Weak vs normal weak eigenstate thermalization} \label{sec:normalwETH}

The few-body observables defined in Sec.~\ref{sec:fewbody} and their sums defined in Sec.~\ref{sec:sums} have $||\hat{O}||^2_\text{mb} = \mathbf{O}(1)$. In contrast, the commonly studied intensive sums of few-body observables $\hat{\mathcal{O}}$, in which one replaces $\sqrt{V}\rightarrow V$ in Eqs.~\eqref{def_T1_T2} and~\eqref{def_O1_O2}, have a norm in the many-body Hilbert space that vanishes in the thermodynamic limit,
\begin{equation} \label{def_norm_inntensive}
    ||\hat{\mathcal{O}}||_\text{mb} \propto \frac{1}{\sqrt{V}}.
\end{equation}
This guarantees that the variance $\sigma^2$ decays at least polynomially with the system size $V$,
\begin{equation}
\label{eq_weak}
    \sigma^2\le ||\hat{O}||_\text{mb}\propto \frac{1}{V},
\end{equation}
where we used Eqs.~(\ref{eq:wETHsums}) and (\ref{def_norm_inntensive}). A simple example of an intensive sum of few-body operators is the average site occupation $\hat{\bar{n}}=\frac{1}{V}\sum_{j}\hat{n}_{j}$, whose norm in the many-body Hilbert space is $||\hat{\bar{n}}||_\text{mb}=1/(2\sqrt{V})$. Hence, regardless of the nature of the studied model, intensive sums of few-body observables exhibit a vanishing variance when calculated over the entire energy spectrum, i.e., they exhibit ``infinite-temperature'' weak eigenstate thermalization.

From Eq.~(\ref{eq_weak}) it follows that, in translationally invariant models, local few-body observables also exhibit ``infinite-temperature'' weak eigenstate thermalization. The diagonal matrix elements of local few-body observables and their corresponding intensive sums in position space are related via
\begin{equation}
    \langle \Omega | \hat{\mathcal{O}} | \Omega \rangle=\frac{1}{V}\sum_{i=1}^{V} \langle \Omega | \hat{o}^{}_i | \Omega \rangle.
\end{equation}
Because of translational invariance, $\langle \Omega | \hat{o}^{}_i | \Omega \rangle$ is independent of $i$, so $\langle \Omega | \hat{\mathcal{O}} | \Omega \rangle = \langle \Omega | \hat{o}^{}_i | \Omega \rangle$, and we see that if $\hat{\mathcal{O}}$ exhibits weak eigenstate thermalization so must $\hat{o}^{}_i$. In this work, the results for few-body observables and for their sums are independent of each other because none of the models considered exhibits translational invariance. 

The previously discussed results open an important question, which is the focus of this work. Namely, are there models for which few-body observables such as those defined in Eqs.~\eqref{eq_o1}--\eqref{eq_o3}, and normalized sums of few-body observables such as those in Eqs.~(\ref{def_T1_T2}) and (\ref{def_O1_O2}), are guaranteed to exhibit  at least polynomially vanishing variances? Namely, are there models for which normalized few-body observables always exhibit normal weak eigenstate thermalization?

\section{Quantum-chaotic vs localized quadratic Hamiltonians} \label{sec:quadraticH}

Central to the analytic derivations and numerical results reported in this work are quadratic Hamiltonians, i.e., Hamiltonians of the form
\begin{equation} \label{deF_H_quadratic}
    \hat H = \sum_{i,j} h_{ij} \hat c_i^\dagger \hat c^{}_j \,,
\end{equation}
where $h_{ij}$ are the Hamiltonian matrix elements in a single-particle basis. Such Hamiltonians can be brought to a diagonal form $\hat H = \sum_{\omega=1}^V \epsilon_\omega \hat f_\omega^\dagger \hat f^{}_\omega$ via a canonical transformation of the fermionic creation and annihilation operators, where $\epsilon_\omega$ are the single-particle eigenenergies. A key property of quadratic Hamiltonians that we will use here is that the many-body eigenstates $|\Omega\rangle$ of $\hat H$ are Slater determinants (fermionic Gaussian states), i.e., they can be written as products of single-particle Hamiltonian eigenstates $|\omega\rangle$.

We report numerical results for two quadratic Hamiltonians that exhibit quantum chaos and eigenstate thermalization in their single-particle spectrum, despite the fact that their matrix elements $h_{ij}$ are not drawn from a fully random ensemble. Our Hamiltonians of interest have a nontrivial structure, and this is the reason behind the fact that quantum chaos only occurs in some parameter regimes and not in others.

We consider the power-law random banded (PLRB) model,
\begin{equation} \label{def_plrb}
    \hat H = \sum_{i,j=1}^V \frac{\mu_{i,j}}{(1 + |i-j|/b)^{a/2}} \hat c_i^\dagger \hat c^{}_j \,,
\end{equation}
where $\mu_{i,j}$ are random variables drawn from a normal distribution with zero mean and unit variance. We take open boundary conditions, and set $a=0.2$ and $b=0.1$ to ensure that the system exhibits single-particle quantum chaos~\cite{mirlin_fyodorov_96, bera_detomasi_18, hopjan_vidmar_23b}.

We also consider the 3D Anderson model on a cubic lattice with $V$ sites:
\begin{equation} \label{def_anderson}
    \hat H = - \sum_{\langle i,j\rangle} \left( \hat c_i^\dagger \hat c^{}_j + {\rm H.c.} \right) + \sum_{i=1}^V \varepsilon_i \, \hat n^{}_i \,,
\end{equation}
where the first sum in Eq.~(\ref{def_anderson}) runs over nearest neighbor sites, and $\varepsilon_i$ are i.i.d.~random variables drawn from a box distribution, $\varepsilon_i \in [-W/2,W/2]$. For this model we take periodic boundary conditions. For nonvanishing disorder strengths below the localization transition, $0 < W < W_{\rm c}$, where $W_{\rm c} \approx 16.5$~\cite{slevin_ohtsuki_18, suntajs_prosen_21}, the 3D Anderson model exhibits quantum chaos in the single-particle sector~\cite{shklovskii_shapiro_93} and single-particle eigenstate thermalization~\cite{lydzba_zhang_21}, i.e., in this regime it is a QCQ model~\cite{lydzba_rigol_21, lydzba_zhang_21}.

We also study quadratic models that do not exhibit quantum chaos nor eigenstate thermalization in the single-particle sector. The single-particle energy eigenstates of such models are localized in position or quasimomentum space, so we refer to them as {\it localized} quadratic models~\cite{lydzba_zhang_21}. We study two such models. The first one is the 3D Anderson model~(\ref{def_anderson}) above the localization transition, $W>W_c$ (we fix $W=30$). The second one is the 1D Aubry–André model
\begin{equation}
    \hat H = - \sum_{i=1}^{V-1} \left( \hat c_i^\dagger \hat c^{}_{i+1} + {\rm H.c.} \right) + \lambda \sum_{i=1}^V \cos(2\pi \eta i+\phi) \hat n^{}_i \,,
\end{equation}
where $\eta = (\sqrt{5}-1)/2$, $\phi$ is randomly drawn from a box distribution, $\phi\in[0,\pi]$, and we take open boundary conditions. The single-particle eigenstates of this model are localized in quasimomentum space for $\lambda<2$, and in position space for $\lambda > 2$~\cite{aubry1980analyticity}. The former regime, being localized in a different basis than the 3D Anderson model at strong disorder, is the focus of our study here. We fix $\lambda=0.5$ in what follows.

In order to evaluate the variance $\sigma^2$ of the diagonal matrix elements across the energy spectrum of the models above, we proceed as follows. For each system size $V$, we first randomly select 200 many-body energy eigenstates at half filling, $n=N/V=1/2$, and calculate the corresponding diagonal matrix elements. We then repeat the calculation for 20 to 100 Hamiltonian realizations. Subsequently, we determine the minimal and maximal energies in the simulations, divide the established energy interval into $100$ bins, and compute the average diagonal matrix elements in each bin. Those averages are then plugged in Eq.~(\ref{def:variance}) to calculate the variance $\sigma^2$ of the given set of matrix elements. [In Eq.~(\ref{def:variance}), we replace $D$ by the corresponding number of matrix elements.]

\section{Normal weak eigenstate thermalization: One-body case} \label{sec:spet_weakETH}

In this section we prove the occurrence of normal weak eigenstate thermalization for one-body observables in QCQ Hamiltonians. We will discuss the extension to two-body observables and beyond in Sec.~\ref{sec:tb_weakETH}.

\subsection{Single-particle eigenstate thermalization} \label{sec:spet}

We begin by emphasizing once again that the normalization introduced in Eq.~(\ref{def_norm_o_mb}) is defined in the many-body Hilbert space, whose dimension $D$ increases exponentially with the number of lattice sites $V$. The normalization of observables is in general different in the single-particle Hilbert space, whose dimension is proportional to $V$. We sharpen that distinction next, as it is instrumental for the derivation of normal weak eigenstate thermalization in QCQ models.

The analog of the ETH ansatz from Eq.~(\ref{def_eth}) was introduced for the matrix elements of one-body observables in the single-particle eigenstates $|\omega\rangle$ of QCQ Hamiltonians in Ref.~\cite{lydzba_zhang_21}. For the diagonal matrix elements $o_{\omega\omega} \equiv \langle \omega |\hat o |\omega\rangle$, which are the ones needed in the derivations that follow, the single-particle eigenstate thermalization ansatz can be written as
\begin{equation}
\label{O_sp}
    \frac{{o}_{\omega\omega}}{||\hat{o}||_\text{sp}}=\oo(\epsilon_\omega)+\frac{1}{\sqrt{\rho(\epsilon_\omega)}}\mathcal{F}_{o}(\epsilon_\omega,0)R_{\omega\omega}^{o},
\end{equation}
where $\rho(\epsilon_\omega)$ is the single-particle density of states, $\oo(\epsilon_\omega)$ and $\mathcal{F}_{o}(\epsilon_\omega,0)$ are smooth functions of the single-particle energies $\epsilon_\omega$, and $R_{\omega\omega}^{o}$ is a random number with zero mean and unit variance.

Two important properties of one-body observables in the single-particle Hilbert space are: (i) One-body observables that are normalized in the many-body Hilbert space, i.e., for which $||\hat{o}||^2_\text{mb} = \mathbf{O}(1)$, exhibit a vanishing norm in the single-particle Hilbert space
\begin{equation} \label{O_str_sp}
    ||\hat{o}||^2_\text{sp} = \mathbf{O}(1/V) \,.
\end{equation}
Consequently, $\sqrt{V}\hat o$ are the observables that are normalized in the single-particle Hilbert-space~\cite{lydzba_zhang_21}. (ii) One-body observables that are normalized in the single-particle Hilbert space have a structure function $o(\epsilon_\omega)$ in Eq.~(\ref{O_sp}) that vanishes in the thermodynamic limit. We prove this property in Appendix~\ref{sec:app_no_structure}.

\subsection{Normal weak eigenstate thermalization} \label{sec:argument}

We are now ready to prove that single-particle eigenstate thermalization implies normal weak eigenstate thermalization for one-body observables in many-body eigenstates. Using Eq.~(\ref{def_oA}), we can write any one-body observable $\hat o$ in terms of the creation \{$\hat{f}_{\omega}^\dagger$\} and annihilation \{$\hat{f}^{}_{\omega}$\} operators of spinless fermions in the single-particle energy eigenstates \{$|\omega\rangle$\}, and the corresponding matrix elements $o_{\omega\gamma}=\langle\omega|\hat{o}|\gamma\rangle$. We focus on one-body observables $\hat o$ that are normalized in the many-body Hilbert-space, namely, for which $||\hat o||^2_\text{mb} = {\bf O}(1)$ [see Eq.~(\ref{def_norm_o_mb})]. The diagonal matrix elements of $\hat{o}$ in many-body energy eigenstates, $o_{\Omega\Omega} = \langle\Omega|\hat o|\Omega\rangle$, can therefore be written as
\begin{equation} \label{O_mb}
    o_{\Omega\Omega} = \sum_{\omega,\gamma=1}^{V} o_{\omega\gamma}\langle\Omega| \hat{f}_{\omega}^\dagger \hat{f}^{}_{\gamma} |\Omega\rangle = \sum_{\omega=1}^{V} o_{\omega\omega}\langle\Omega| \hat{f}_{\omega}^\dagger \hat{f}^{}_{\omega} |\Omega\rangle.
\end{equation}
Plugging the single-particle eigenstate thermalization ansatz for $o_{\omega\omega}$ [see Eq.~(\ref{O_sp})] in Eq.~(\ref{O_mb}) results in two contributions to $o_{\Omega\Omega}$,
\begin{equation}
    o_{\Omega\Omega} = \chi_{1,\Omega} + \chi_{2,\Omega} \,,
\end{equation}
of the form
\begin{align}
    \chi_{1,\Omega} & = ||\hat{o}||_\text{sp} \sum_{\omega=1}^{V}\oo(\epsilon_\omega)\langle\Omega|\hat{f}^\dagger_\omega\hat{f}^{}_{\omega}|\Omega\rangle \label{def_chi_1}\,, \\
    \chi_{2,\Omega} & = ||\hat{o}||_\text{sp} \sum_{\omega=1}^{V}\frac{1}{\sqrt{\rho(\epsilon_\omega)}}\mathcal{F}_{o}(\epsilon_\omega,0)R_{\omega\omega}^{o}\langle\Omega|\hat{f}^\dagger_\omega\hat{f}^{}_{\omega}|\Omega\rangle.
    \label{def_chi_2}
\end{align}

The eigenstate fluctuations of $o_{\Omega\Omega}$ due to $\chi_{2,\Omega}$ were recently studied in Ref.~\cite{lydzba_mierzejewski_23}. The main result of that study is that $\chi_{2,\Omega}$ yields no structure in the many-body sector. In other words, the microcanonical average of $\chi_{2,\Omega}$ is energy independent. Using this result, we are able to calculate the variance of $\chi_{2,\Omega}$ analytically, as detailed in Appendix~\ref{app:variance_chi_2}. We obtain
\begin{equation} \label{def_variance_chi2a}
    {\rm Var}[\chi_{2,\Omega}] = V ||\hat o||_{\rm sp}^2 \, n (1-n)\, \overline {A^2}_{\rm sp} \,,
\end{equation}
where $n$ is the average particle filling in the many-body Hilbert space, and
\begin{equation} \label{def_A2_sp}
    \overline {A^2}_{\rm sp} \equiv \frac{1}{V} \sum_{\omega=1}^V \frac{1}{\rho(\epsilon_\omega)} |\mathcal{F}_{o}(\epsilon_\omega,0)|^2 |R_{\omega\omega}^{o}|^2 \,.
\end{equation}
The sum in Eq.~\eqref{def_A2_sp} has $V$ terms that are ${\bf O}(1/V)$, so it is ${\bf O}(1)$, which leads to $\overline {A^2}_{\rm sp} \propto 1/V$ because of the prefactor of the sum. Recalling that $V ||\hat o||_{\rm sp}^2 = {\bf O}(1)$, see Eq.~(\ref{O_str_sp}), we find that
\begin{equation} \label{def_variance_chi2}
    {\rm Var}[\chi_{2,\Omega}] \propto \frac{1}{V}.
\end{equation}
While ${\rm Var}[\chi_{2,\Omega}]$ vanishes polynomially with increasing $V$, we should stress that in Ref.~\cite{lydzba_mierzejewski_23} we proved that there is an exponentially large (in $V$) number of matrix elements $\chi_{2,\Omega}$ that are different from the average as $V\rightarrow\infty$.

Next, we study the contribution of $\chi_{1,\Omega}$ in Eq.~(\ref{def_chi_1}). Following Ref.~\cite{mierzejewski_vidmar_20}, we express the structure function ${\cal O}(\epsilon_\omega)$ as a polynomial, 
\begin{equation} \label{def_structure_polynomial}
    \oo(\epsilon_\omega)=\sum_{i=0}^{V} p_{i}(\epsilon_\omega)^i\,,
\end{equation}
where $\epsilon_\omega$ are the single-particle energy eigenvalues and $p_i$ are coefficients that characterize each observable, see Appendix~\ref{sec:app_no_structure}. In Appendix~\ref{sec:app_no_structure}, we prove that for one-body observables the coefficients $p_i$ are at most ${\bf O}(1/\sqrt{V})$, while for sums of one-body observables they can be ${\bf O}(1)$.

Using Eq.~(\ref{def_structure_polynomial}), one can write $\chi_{1,\Omega}$ as
\begin{align}
\chi_{1,\Omega} = &\, ||\hat{o}||_\text{sp}\sum_{i=0}^{V}p_i\sum_{\omega=1}^{V}(\epsilon_\omega)^i\langle\Omega|\hat{f}^\dagger_\omega\hat{f}_\omega|\Omega\rangle \\
= &\, ||\hat{o}||_\text{sp}\,p_0N + ||\hat{o}||_\text{sp}\,p_1E_\Omega \label{O_st} \nonumber \\ 
&\, + ||\hat{o}||_\text{sp}\sum_{i=2}^{V}p_i\sum_{\omega=1}^{V}(\epsilon_\omega)^i\langle\Omega|\hat{f}^\dagger_\omega\hat{f}_\omega|\Omega\rangle\,. 
\end{align}
The first and second terms on the r.h.s.~of Eq.~(\ref{O_st}) have a simple interpretation: the first term is a constant shift and the second term introduces a linear ($\propto E_\Omega$) structure in the diagonal matrix elements $o_{\Omega\Omega}$ in the many-body energy eigenstates. The third term in Eq.~(\ref{O_st}) introduces further structure, and it is the first term that introduces eigenstate to eigenstate fluctuations of $o_{\Omega\Omega}$. These fluctuations, as argued next, can result in a nonvanishing variance of the diagonal matrix elements of normalized sums of one-body observables.

\begin{figure}[t!]
\includegraphics[width=0.98\columnwidth]{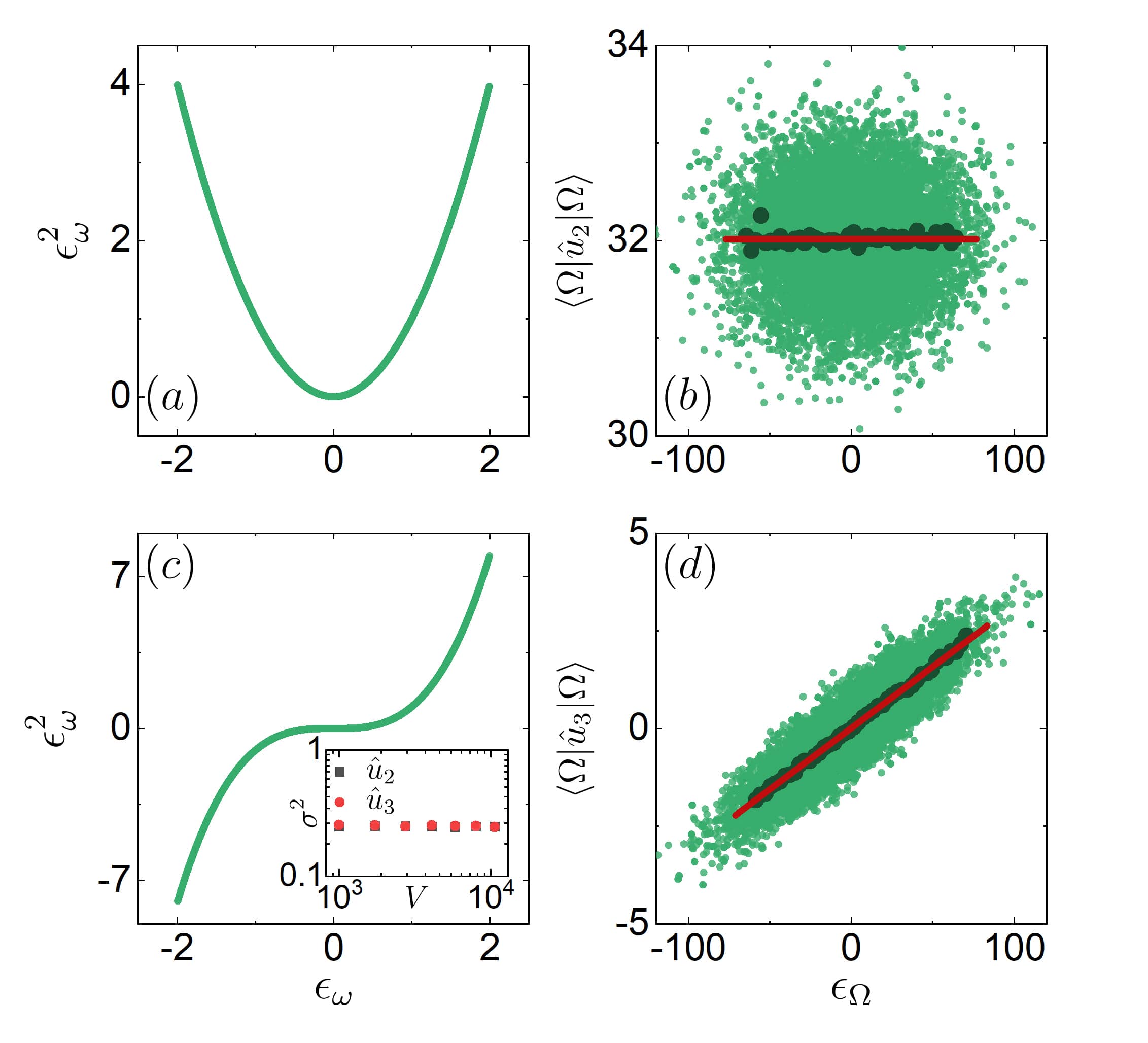}
\vspace{-0.2cm}
\caption{Diagonal matrix elements of the observables $\hat{\rm u}_p$ in Eq.~\eqref{eq:upobservables} for (a),(b) $p=2$ and (c),(d) $p=3$ in (a),(c) single-particle and (b),(d) many-body eigenstates of the SYK2 model with $V=4096$ sites. The calculations for the many-body eigenstates are carried out at half filling, i.e., for $N=V/2$. In (a),(c), we plot all the diagonal matrix elements for two realizations of the Hamiltonian. In (b),(d), we plot $200$ diagonal matrix elements for $60$ realizations of the Hamiltonian. The dark green points show bin averages of the light green points computed dividing the energy range shown in 100 identical bins. The red curves show least-square fits of the dark points to (a) a constant and (d) a linear function. The inset in (c) displays the variance $\sigma^2$ [see Eq.~(\ref{def:variance})] of $\hat{u}_p$ for $p=2$ (grey squares) and $p=3$ (red circles).}
\label{figa1}
\end{figure}

To explore the possible behaviors of $\chi_{1,\Omega}$ numerically, we use the quadratic Sachdev-Ye-Kitaev (SYK2) model~\cite{lydzba_rigol_20, lydzba_rigol_21, lydzba_zhang_21}, in which the matrix elements in Eq.~(\ref{deF_H_quadratic}) are drawn from the Gaussian orthogonal ensemble, and we consider the following ($p$-dependent) observables
\begin{equation}\label{eq:upobservables}
    \hat {\rm u}_p = \frac{1}{\sqrt{V}}\sum_{\omega=1}^V (\epsilon_\omega)^p \hat f_\omega^\dagger \hat f^{}_\omega \,.
\end{equation}
The matrix elements of $\hat {\rm u}_p$ in the single-particle energy eigenstates are smooth by construction, i.e., the fluctuating part of the single-particle eigenstate thermalization ansatz in Eq.~(\ref{O_sp}) vanishes for any system size. This is illustrated for $\hat {\rm u}_2$ and $\hat {\rm u}_3$ in Figs.~\ref{figa1}(a) and~\ref{figa1}(c), respectively. Hence, the only contribution to the diagonal matrix elements of $\hat {\rm u}_p$ in the many-body eigenstates comes from $\chi_{1,\Omega}$ in Eq.~(\ref{O_st}).

In Figs.~\ref{figa1}(b) and~\ref{figa1}(d), we show the matrix elements of $\hat {\rm u}_2$ and $\hat {\rm u}_3$ in the many-body eigenstates. In addition to the eigenstate to eigenstate fluctuations, there is no structure for $\hat {\rm u}_2$ [Fig.~\ref{figa1}(b)], while there is structure for $\hat {\rm u}_3$ [Fig.~\ref{figa1}(d)]. This shows that the structure in the diagonal matrix elements in the single-particle sector may or may not result in structure of the diagonal matrix elements in many-body eigenstates. We find similar results (not shown), in which there is no structure for ${\rm u}_p$ when $p$ is even, and there is a structure for ${\rm u}_p$ when $p$ is odd, for larger values of $p$.

With this knowledge in hand, we study the scaling of the variance of $\chi_{1,\Omega}$. We note that, because of the structure in $\chi_{1,\Omega}$, calculating the scaling of the variance of $\chi_{1,\Omega}$ is more challenging than calculating the one of $\chi_{2,\Omega}$. The reason is that we are interested in the fluctuations of the matrix elements about the structure. Since we do not know the structure a priori, we provide an upper bound for the fluctuations by calculating the variance of the terms with $i\geq 2$ in Eq.~(\ref{O_st}),
\begin{equation} \label{def_chi_1_omega_i}
    \chi_{1,\Omega}^{(i)} = ||\hat o||_{\rm sp}\, p_i \sum_{\omega=1}^{V}(\epsilon_\omega)^i\langle\Omega|\hat{f}^\dagger_\omega\hat{f}^{}_\omega|\Omega\rangle\,.
\end{equation} 
We find that, see Appendix~\ref{app:variance_chi_1},
\begin{equation} \label{def_variance_chi1}
    {\rm Var}[\chi_{1,\Omega}^{(i)}] = V||\hat{o}||^2_\text{sp}\, n(1-n)\, p_i^2\, \overline{\epsilon^{2i}}_{\rm sp} \,,\,\,\, i\geq 2 \,,
\end{equation}
where $\overline{\epsilon^{2i}}_{\rm sp}= {\bf O}(1)$ is the mean value of the single-particle energies $(\epsilon_\omega)^{2i}$ in the single-particle sector, see Eq.~(\ref{def_epsilon_avr}). Also, $V ||\hat o||_{\rm sp}^2 = {\bf O}(1)$, see Eq.~(\ref{O_str_sp}).

Equation~\eqref{def_variance_chi1} shows that if $p_i = {\bf O}(1/\sqrt{V})$, which is the case for one-body observables, ${\rm Var}[\chi_{1,\Omega}^{(i)}] \propto 1/V$. The latter scaling is the same as the one in Eq.~(\ref{def_variance_chi2}). On the other hand, if $p_i = {\bf O}(1)$, which is the case for sums of one-body observables, ${\rm Var}[\chi_{1,\Omega}^{(i)}] = {\bf O}(1)$.

For even $i$ (with $i\geq 2$), for which there is no structure in the diagonal matrix elements $o_{\Omega\Omega}$, ${\rm Var}[\chi_{1,\Omega}^{(i)}]$ in Eq.~(\ref{def_variance_chi1}) gives the scaling of the fluctuations of the diagonal matrix elements $o_{\Omega\Omega}$. For odd $i$ (with $i \geq 3$), for which there is a structure in diagonal matrix elements $o_{\Omega\Omega}$, ${\rm Var}[\chi_{1,\Omega}^{(i)}]$ only provides an upper bound for the scaling of the fluctuations of the diagonal matrix elements $o_{\Omega\Omega}$. Therefore, for (normalized) sums of one-body observables [for which $p_i = {\bf O}(1)$], Eq.~(\ref{def_variance_chi1}) shows that normal weak eigenstate thermalization may or may not occur. In the inset of Fig.~\ref{figa1}(c), we report our numerical results for $\sigma^2$ [see Eq.~\eqref{def:variance}] of the observables $\hat{\rm u}_{p}$ from Eq.~\eqref{eq:upobservables} for $p=2$ (grey squares) and $p=3$ (red circles). The variances were calculated from $200$ eigenstates and averaged over $60$ Hamiltonian realizations. In agreement with our analytical calculations, we find that $\sigma^2$ is ${\bf O}(1)$ for those two observables, i.e., they do not exhibit normal weak eigenstate thermalization.

Summarizing our analysis in this section, we have proved that, for one-body observables, normal weak eigenstate thermalization always occurs in QCQ Hamiltonians. On the other hand, for sums of one-body observables, normal weak eigenstate thermalization may or may not occur in QCQ Hamiltonians.

\section{Numerical results: One-body case} \label{sec:spet_weakETH_numerical}

In this section we test our analytical predictions using numerical calculations to solve quadratic models that exhibit quantum chaos in the single-particle sector (QCQ models), and contrast the results to those obtained in models that do not exhibit single particle quantum chaos (localized quadratic models).

\subsection{Quantum-chaotic quadratic Hamiltonians} \label{sec:chaotic}

\begin{figure}[!t]
\includegraphics[width=0.98\columnwidth]{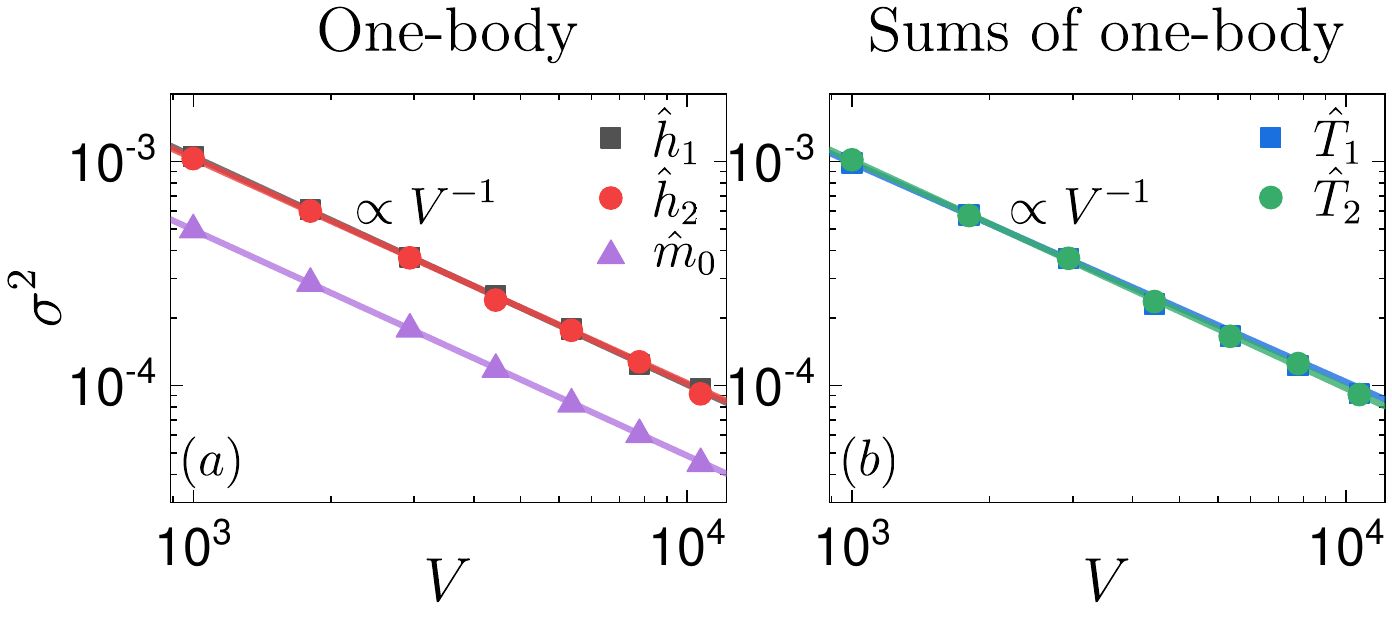}
\vspace{-0.2cm}
\caption{Variances $\sigma^2$, see Eq.~(\ref{def_plrb}), as functions of $V$ for the PLRB model with $a=0.2$ and $b=0.1$. (a) One-body observables, (b) sums of one-body observables. The straight lines show two-parameter fits $aV^{-b}$ to the results. We calculate $\sigma^2$ using $200$ random eigenstates from each of $100$ ($20$) Hamiltonian realizations for $V\le16^3$ ($V>16^3$).}
\label{fig_plrb}
\end{figure}

First, we calculate the variances of one-body observables and their sums in the PLRM model. Results for those observables are shown in Fig.~\ref{fig_plrb}. As predicted analytically, for the one-body observables $\hat h_1$, $\hat h_2$, and $\hat m_0$ [Fig.~\ref{fig_plrb}(a)], we find that $\sigma^2 \propto 1/V$, i.e., they exhibit normal weak eigenstate thermalization. For this model, because of the absence of structure for $\hat T_1$ and $\hat T_2$ in the single-particle sector, we find that those sums of few-body observables also exhibit normal weak eigenstate thermalization [$\sigma^2 \propto 1/V$, see Fig.~\ref{fig_plrb}(b)].

\begin{figure}[!b]
\includegraphics[width=0.98\columnwidth]{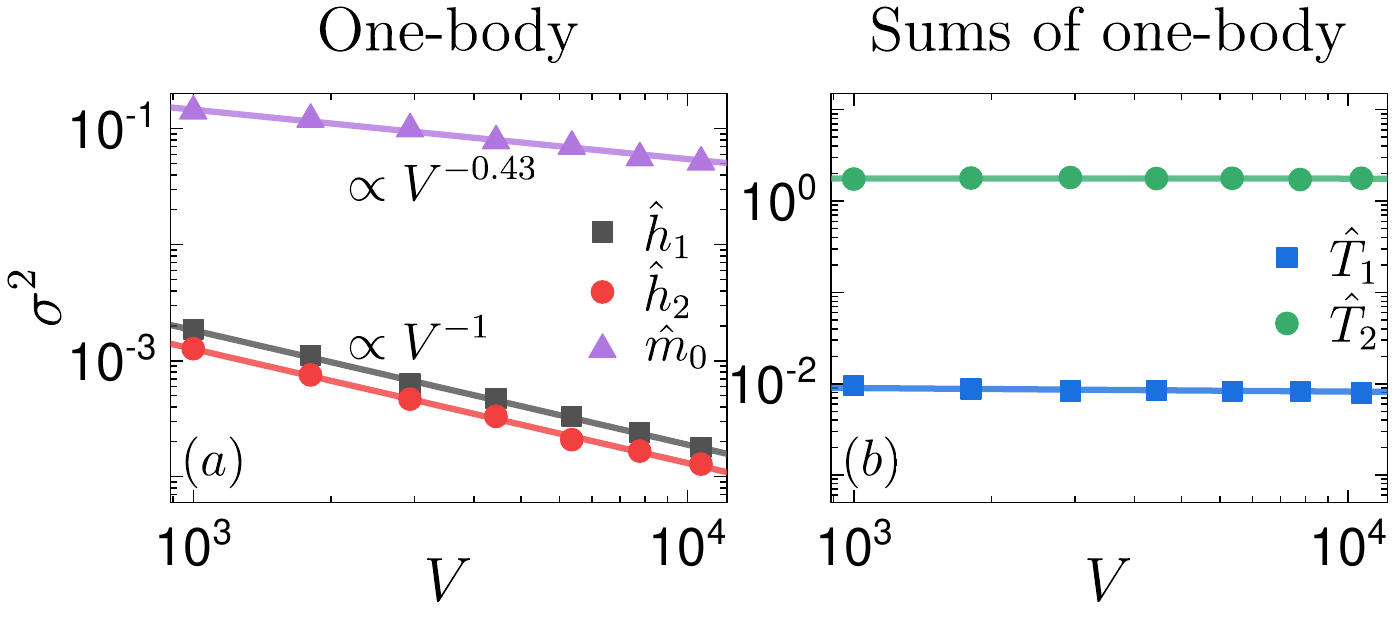}
\vspace{-0.2cm}
\caption{Same as Fig.~\ref{fig_plrb} in the delocalized regime, $W=5$, of the 3D Anderson model.}
\label{fig_anderson_1}
\end{figure}

Next, we calculate the variances of one-body observables and their sums in the delocalized regime of the 3D Anderson model for $W=5$, i.e., in the quantum chaotic regime of the single-particle sector. The results are shown in Fig.~\ref{fig_anderson_1}. For the one-body observables, see Fig.~\ref{fig_anderson_1}(a), we find that $\hat h_1$ and $\hat h_2$ exhibit the normal weak eigenstate thermalization scaling, $\sigma^2 \propto 1/V$. The occupation of the zero quasimomentum mode $\hat m_0$, on the other hand, exhibits a slower decay. We expect this to be a finite-size effect that is a consequence of the proximity to the translationally invariant point at $W=0$. All single-particle energy eigenstates at $W=0$ are localized in quasimomentum space (they are Bloch states). Full delocalization in quasimomentum space at $W=5$, so that $\sigma^2 \propto 1/V$ as predicted analytically, requires larger system sizes that those accessible to us here. 

For the sums of one-body observables, see Fig.~\ref{fig_anderson_1}(b), we find that $\hat T_1$ and $\hat T_2$ do not exhibit normal weak eigenstate thermalization as their variances do not decrease with increasing $V$. This occurs because of the structure in the single-particle sector, which for $\hat T_1$ was studied in Ref.~\cite{lydzba_zhang_21}. We note that $\sigma^2$  for $\hat T_1$ is much smaller than for $\hat T_2$. This is a consequence of the diagonal matrix elements of $\hat T_1$ in the single-particle sector being mostly proportional to the single-particle energies [i.e., $p_1>0$ in Eq.~(\ref{O_st})]. The deviations from this linear behavior are the ones responsible for the breakdown of normal weak eigenstate thermalization. On the other hand, the diagonal matrix elements of $\hat T_2$ in the single-particle sector are mostly proportional to the square of the single-particle energies, which directly results in the breakdown of normal weak eigenstate thermalization.

\begin{figure}[!t]
\includegraphics[width=0.95\columnwidth]{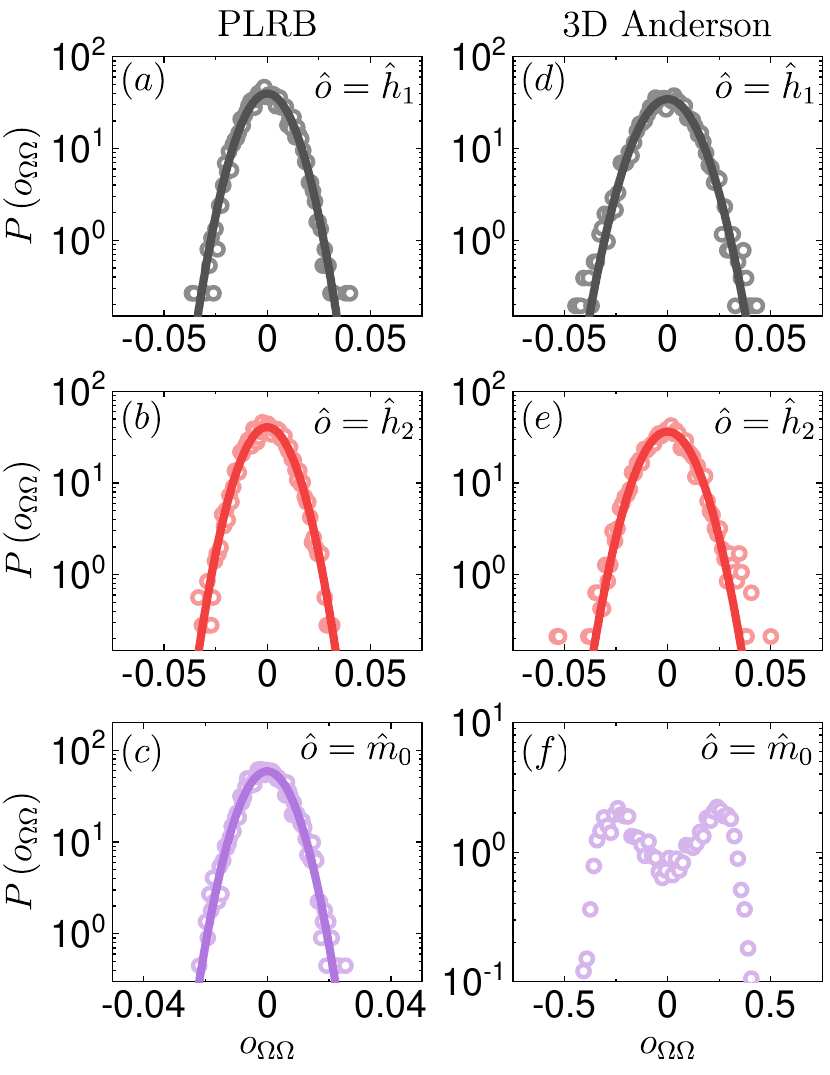}
\vspace{-0.2cm}
\caption{Distributions of the diagonal matrix elements of (a),(d) $\hat{h}_{1}$, (b),(e) $\hat{h}_{2}$, and (c),(f) $\hat{m}_{0}$ in (a)--(c) the PRBM model with $\alpha=0.2$ and $\beta=0.1$, and (d)--(f) the 3D Anderson model with $W=5$. We consider $V=10648$ ($V=22^3=10648$) lattice sites and $200$ diagonal matrix elements for $20$ realizations of the PRBM (3D Anderson) model. The lines show least-squares fits to a Gaussian function.}
\label{fig_1body_dist}
\end{figure}

We also study the distributions of the diagonal matrix elements of one-body observables. Note that, in Eq.~(\ref{O_mb}), the matrix elements in the many-body eigenstates are written as weighted sums of the matrix elements in the single-particle eigenstates. Hence, given that the single-particle sector exhibits quantum chaos, for sufficiently large system sizes one expects the matrix elements in the many-body eigenstates to be normally distributed. This is the case, as in Ref.~\cite{zhang_vidmar_22}, for most of the observables under investigation for the system sizes considered.

Specifically, in Fig.~\ref{fig_1body_dist} we plot the probability density functions (PDFs) of the diagonal matrix elements of one-body observables in the PLRB model (with $\alpha=0.2$ and $\beta=0.1$, left column) and in the 3D Anderson model (with $W=5$, right column), respectively. For all but one observable in one model, the distributions are close to Gaussian as advanced. The exception is $\hat m_0$ in the 3D Anderson model, see Fig.~\ref{fig_1body_dist}(e), which exhibits a bimodal distribution. The corresponding matrix elements are shown in Fig.~\ref{fig_anderson_m0} in Appendix~\ref{app:distributions}. The results for $\hat m_0$ make apparent that, in our finite systems, the matrix elements tend to accumulate at two values that depart from the mean. This is a remnant of what happens for $W=0$, for which the matrix elements of $\hat m_0$ are either 0 or 1 depending on whether the single-particle ground state is included or not in the many-body eigenstate. On the other hand, the PDF of the diagonal matrix elements of $\hat m_0$ in the PLRB model, see Fig.~\ref{fig_1body_dist}(e), is very close to Gaussian. We expect that, for sufficiently large systems (after full delocalization occurs in quasimomentum space), the matrix elements of $\hat m_0$ in the 3D Anderson model with $W=5$ will be normally distributed about the mean value as for all other observables shown.

\subsection{Localized quadratic Hamiltonians} \label{sec:localized}

\begin{figure}[!b]
\includegraphics[width=0.98\columnwidth]{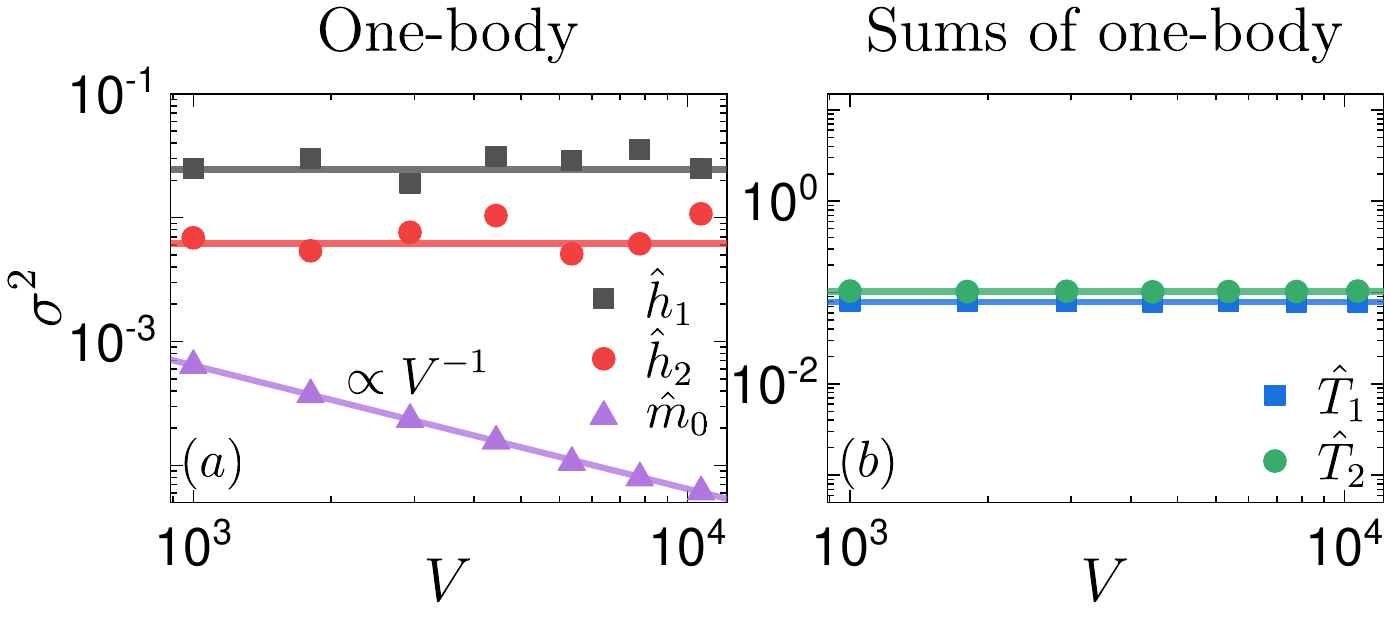}
\vspace{-0.2cm}
\caption{Variances $\sigma^2$, see Eq.~(\ref{def_plrb}), as functions of $V$ for the 3D Anderson model with $W=30$. (a) One-body observables, (b) sums of one-body observables. The straight lines show two-parameter fits $aV^{-b}$ to the results. We calculate $\sigma^2$ using $200$ random eigenstates from each of $100$ ($20$) Hamiltonian realizations for $V\le16^3$ ($V>16^3$).}
\label{fig_anderson_2}
\end{figure}

We showed in previous sections that eigenstate thermalization in the single-particle sector results in normal weak eigenstate thermalization of one-body observables in many-body eigenstates of quadratic models. Our main conjecture for models that do not exhibit single-particle quantum chaos (because of single-particle localization) is that the lack of eigenstate thermalization in the single-particle sector results in no normal weak eigenstate thermalization of one-body observables, specifically, of one-body observables that are local in the basis in which localization occurs in the single-particle sector.

We first test this conjecture for the 3D Anderson model with $W=30$, see Fig.~\ref{fig_anderson_2}. For this model, in which single-particle energy eigenstates are localized in position space, normal weak eigenstate thermalization does not occur for observables that are local in position space~\cite{lydzba_zhang_21}. Figure~\ref{fig_anderson_2}(a) shows that the variance $\sigma^2$ of the diagonal matrix elements of $\hat{h}_1$ and $\hat{h}_2$ in the many-body energy eigenstates does not decay with $V$, i.e., they do not exhibit normal weak eigenstate thermalization. In contrast, the occupation of the zero quasimomentum mode $\hat m_0$ exhibits normal weak eigenstate thermalization ($\sigma^2 \propto 1/V$). In Ref.~\cite{lydzba_zhang_21}, $\hat m_0$ was found to exhibit signatures of single-particle eigenstate thermalization in the localized regime. Figure~\ref{fig_anderson_2}(b) shows that $\hat T_1$ and $\hat T_2$, which are sums of local one-body observables, do not exhibit normal weak eigenstate thermalization [$\sigma^2={\bf O}(1)$].

\begin{figure}[!t]
\includegraphics[width=0.98\columnwidth]{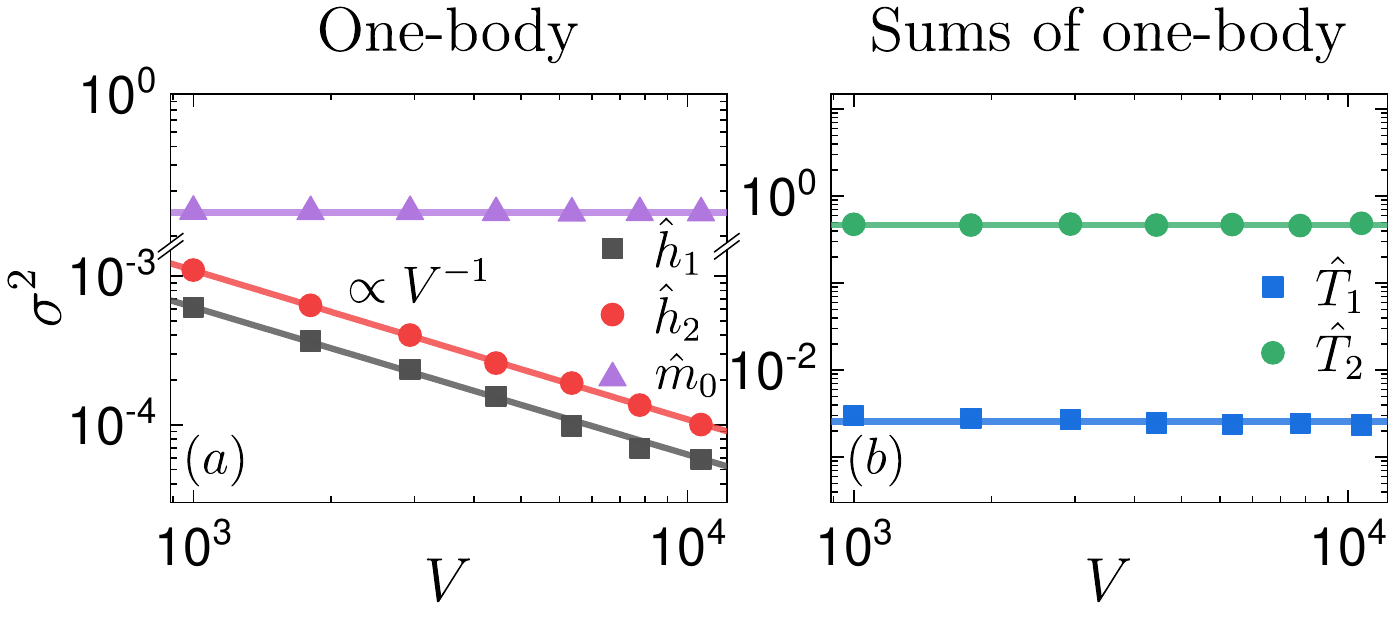}
\vspace{-0.2cm}
\caption{Same as Fig.~\ref{fig_anderson_2} for the Aubry–André model with $\lambda=0.5$, for which single-particle eigenstates are delocalizad in position space and localized in quasimomentum space.}
\label{fig_aa}
\end{figure}

For the 1D Aubry–André model with $\lambda=0.5$, for which single-particle eigenstates are delocalizad in position space and localized in quasimomentum space, the scaling of the variances of one-body observables is opposite to that of the 3D Anderson model localized in position space. Specifically, Fig.~\ref{fig_aa}(a) shows that the variance of both $\hat h_1$ and $\hat h_2$ scales as $\sigma^2 \propto 1/V$, while the variance of $\hat m_0$ does not decay with $V$. On the other hand, $\hat T_1$ and $\hat T_2$ (which are sums of local one-body observables) do not exhibit normal weak eigenstate thermalization in Fig.~\ref{fig_aa}(b). The results in Figs.~\ref{fig_anderson_2}(b) and~\ref{fig_aa}(b) show that the lack of normal weak eigenstate thermalization for sums of one-body observables in these models is independent of whether the observables are local or not in the basis in which the single-particle energy eigenstates are localized.

\section{Normal weak eigenstate thermalization: Two-body case} \label{sec:tb_weakETH}

In this section, we provide analytical arguments and report numerical results to argue that, like one-body observables, two-body observables always exhibit normal weak eigenstate thermalization in QCQ models, i.e., a variance $\sigma^2\le\mathbf{O}(1/V)$. A detailed analytical calculation for observables that are products of occupation operators can be found in Appendix~\ref{app:2body_arguments}. 

\subsection{Analytical arguments} \label{sec:beyond_onebody}

Let us consider a {\it more general} two-body observable $\hat{a}_{1}\hat{a}_{2}\hat{a}_{3}\hat{a}_{4}$, where $\hat a_j$ is either a fermionic creation or annihilation operator in an arbitrary single-particle state $|j\rangle$. The corresponding matrix elements in the many-body eigenstates can be written as $(a_1 a_2 a_3 a_4)_{\Omega\Omega} = \langle \Omega|\hat{a}_{1}\hat{a}_{2}\hat{a}_{3}\hat{a}_{4}|\Omega\rangle$. For quadratic models, the latter can be related to the expectation values of one-body observables using Wick's decomposition~\cite{Kita2015},
\begin{equation} \label{def_wick}
    (a_{1}a_{2}a_{3}a_{4})_{\Omega\Omega} = \sum_{P\in S_4}\!{}^{'} \text{sgn}(P)\, (a_{P_1}a_{P_2})_{\Omega\Omega}\, (a_{P_3}a_{P_4})_{\Omega\Omega}\,,
\end{equation}
where $S_4$ is the permutation group of $4$ elements and the sum is restricted (hence the prime) to the permutations with $P_1 < P_2$ and $P_3 < P_4$, as well as $P_1 < P_3$. The first and second conditions maintain the alignment order, while the third condition excludes double counting. 

To calculate the variance of $(a_{1}a_{2}a_{3}a_{4})_{\Omega\Omega}$, we carry out two further simplifications. First, since $(a_{1}a_{2}a_{3}a_{4})_{\Omega\Omega}$ in Eq.~(\ref{def_wick}) is expressed as a sum of different contributions, one can use the covariance inequality (based on the Cauchy-Schwarz inequality) to upper bound the variance:
\begin{equation} \label{covariance_inequality}
    \sigma^2[(a_{1}a_{2}a_{3}a_{4})_{\Omega\Omega}]\! \le \!\left(\sum_{P\in S_4}\!{}^{'}\,\sigma\left[
    (a_{P_1}a_{P_2})_{\Omega\Omega} (a_{P_3}a_{P_4})_{\Omega\Omega}
    \right]\right)^2\!\!.
\end{equation}
Second, the variances of products of matrix elements of two different observables, which appear on the r.h.s.~of the inequality in Eq.~(\ref{covariance_inequality}), can be expressed as
\begin{eqnarray} \label{def_variance_product}
    &&\sigma^2 \left[ (a_{P_1}a_{P_2})_{\Omega\Omega} (a_{P_3}a_{P_4})_{\Omega\Omega} \right]\nonumber\\ &&\quad= \sigma^2[(a_{P_1} a_{P_2})_{\Omega\Omega}] \, \sigma^2[(a_{P_3} a_{P_4})_{\Omega\Omega}] \nonumber \\ &&\qquad+\sigma^2[(a_{P_1} a_{P_2})_{\Omega\Omega}] \,\mathbb{E}^2[(a_{P_3} a_{P_4})_{\Omega\Omega}] \nonumber\\&&\qquad+\sigma^2[(a_{P_3} a_{P_4})_{\Omega\Omega}] \, \mathbb{E}^2[(a_{P_1} a_{P_2})_{\Omega\Omega}]
    \,,
\end{eqnarray}
where we assumed a vanishing covariance between different matrix elements (independent matrix elements) and $\mathbb{E}[...]$ is the mean defined over the same set of states as the variance $\sigma^2$. Assuming that the one-body observables exhibit single-particle eigenstate thermalization, and using that (as we proved) the variances of one-body observables in such case scale as $\sigma^2 = {\bf O}(1/V)$, we find the scaling of the variance in Eq.~(\ref{def_variance_product}) to be
\begin{align}
    & \sigma^2 \left[ (a_{P_1}a_{P_2})_{\Omega\Omega} (a_{P_3}a_{P_4})_{\Omega\Omega} \right]
    =\mathbf{O}(1/V^2) \label{twbody_scalings} \\
    & +\mathbf{O}(1/V)\, \mathbb{E}^2[(a_{P_3} a_{P_4})_{\Omega\Omega}] +\mathbf{O}(1/V)\, \mathbb{E}^2[(a_{P_1} a_{P_2})_{\Omega\Omega}] \,. \nonumber
\end{align}
We therefore see that when the means $\mathbb{E}[...]$ vanish:
\begin{equation} \label{twobody_scenario_1}
     \sigma^2[(a_{1}a_{2}a_{3}a_{4})_{\Omega\Omega}] \le \mathbf{O}(1/V^2)\,.
\end{equation}
We show in Sec.~\ref{sec:2body_numerics} that this is what happens for $\hat o_3$ in Eq.~(\ref{eq_o3}). On the other hand, when the means $\mathbb{E}[...]$ do not vanish, which means that they are $\mathbf{O}(1)$, we get
\begin{equation} \label{twobody_scenario_2}
     \sigma^2[(a_{1}a_{2}a_{3}a_{4})_{\Omega\Omega}] \le \mathbf{O}(1/V)\,.
\end{equation}
We show in Sec.~\ref{sec:2body_numerics} that this is what happens for $\hat o_1$ and $\hat o_2$ from Eqs.~(\ref{eq_o1}) and (\ref{eq_o2}), respectively. For the latter observables, we provide an explicit derivation of Eq.~(\ref{twobody_scenario_2}) in Appendix~\ref{app:2body_arguments}.

No matter which of the two scalings applies to any given two-body observable, Eq.~(\ref{twobody_scenario_1}) or~(\ref{twobody_scenario_2}), we say that it exhibits normal weak eigenstate thermalization as the variance vanishes polynomially with increasing $V$. 

The previous calculations can be extended to $p$-body observables with $p>2$, provided that $p={\bf O} (1)$. When $p={\bf O} (V)$, the upper bound for the variance from Eq.~(\ref{covariance_inequality}) may no longer decay polynomially with the system size. For sums of two-body observables, we provide numerical evidence that normal weak eigenstate thermalization occurs for the observables considered. Its analytical explanation is something we plan to explore in future works, together with what happens for sums of three- and higher few-body observables. 

\subsection{Numerical results} \label{sec:2body_numerics}

Next, we show numerical results for two-body observables, and sums of two-body observables, in the 3D Anderson, the PLRB, and the 1D Aubry–André models.

\begin{figure}[!b]
\includegraphics[width=0.98\columnwidth]{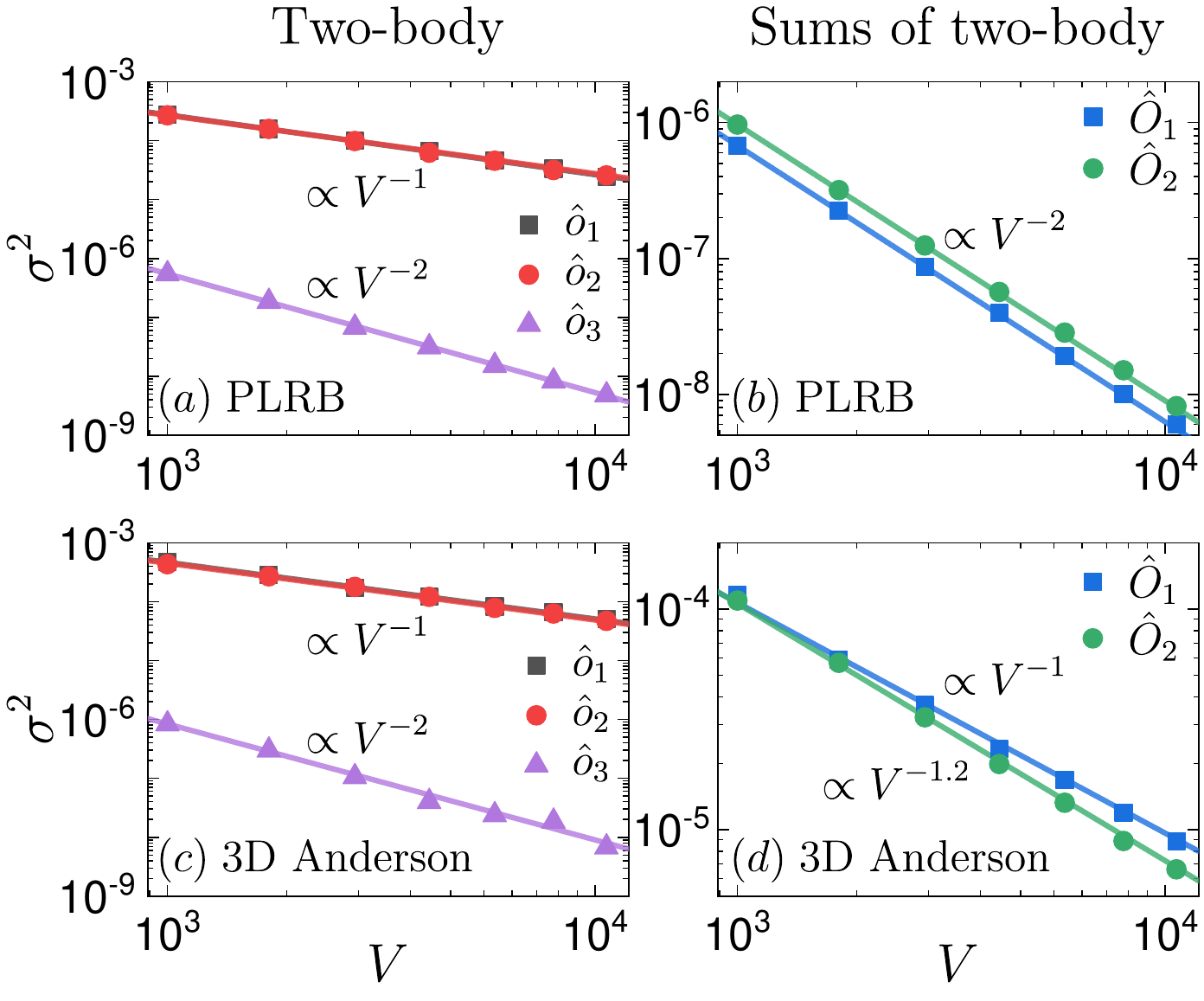}
\vspace{-0.2cm}
\caption{Variances $\sigma^2$, see Eq.~(\ref{def_plrb}), as functions of $V$ for the (a),(b) PLRB model with $\alpha=0.2$, $\beta=0.1$ and (c),(d) 3D Anderson model with $W=5$. (a),(c) Two-body observables, (b),(d) sums of two-body observables. The straight lines show two-parameter fits $aV^{-b}$ to the results. We calculate $\sigma^2$ using $200$ random eigenstates from each of $100$ ($20$) Hamiltonian realizations for $V\le16^3$ ($V>16^3$).}
\label{fig_anderson_plrb_twobody}
\end{figure}

In Fig.~\ref{fig_anderson_plrb_twobody}, we show results for the variances $\sigma^2$ in the (single-particle) quantum-chaotic regime of the PLRM model with $\alpha=0.2$ and $\beta=0.1$ [Figs.~\ref{fig_anderson_plrb_twobody}(a) and~\ref{fig_anderson_plrb_twobody}(b)] and of the 3D Anderson model with $W=5$ [Figs.~\ref{fig_anderson_plrb_twobody}(c) and~\ref{fig_anderson_plrb_twobody}(d)]. The results for two-body observables are shown in Figs.~\ref{fig_anderson_plrb_twobody}(a) and~\ref{fig_anderson_plrb_twobody}(c) and for sums of two-body observables in Figs.~\ref{fig_anderson_plrb_twobody}(b) and~\ref{fig_anderson_plrb_twobody}(d). All those observables exhibit normal weak eigenstate thermalization. However, the scaling of the variance depends on the observable. In Figs.~\ref{fig_anderson_plrb_twobody}(a) and~\ref{fig_anderson_plrb_twobody}(c), the variances of $\hat o_1$ and $\hat o_2$ scale as $\sigma^2\propto 1/V$, while the variance of $\hat o_3$ scales as $\sigma^2\propto 1/V^2$. Both scalings were argued in Sec.~\ref{sec:beyond_onebody} to occur for two-body observables. Figures~\ref{fig_anderson_plrb_twobody}(b) and~\ref{fig_anderson_plrb_twobody}(d) show that the sums of two body observables $\hat O_1$ and $\hat O_2$, exhibit a scaling that is close to $\sigma^2\propto 1/V^\zeta$ with $\zeta=1$ or $2$ depending on the model. Why those scalings emerge is yet to be understood analytically.

\begin{figure}[!t]
\includegraphics[width=0.98\columnwidth]{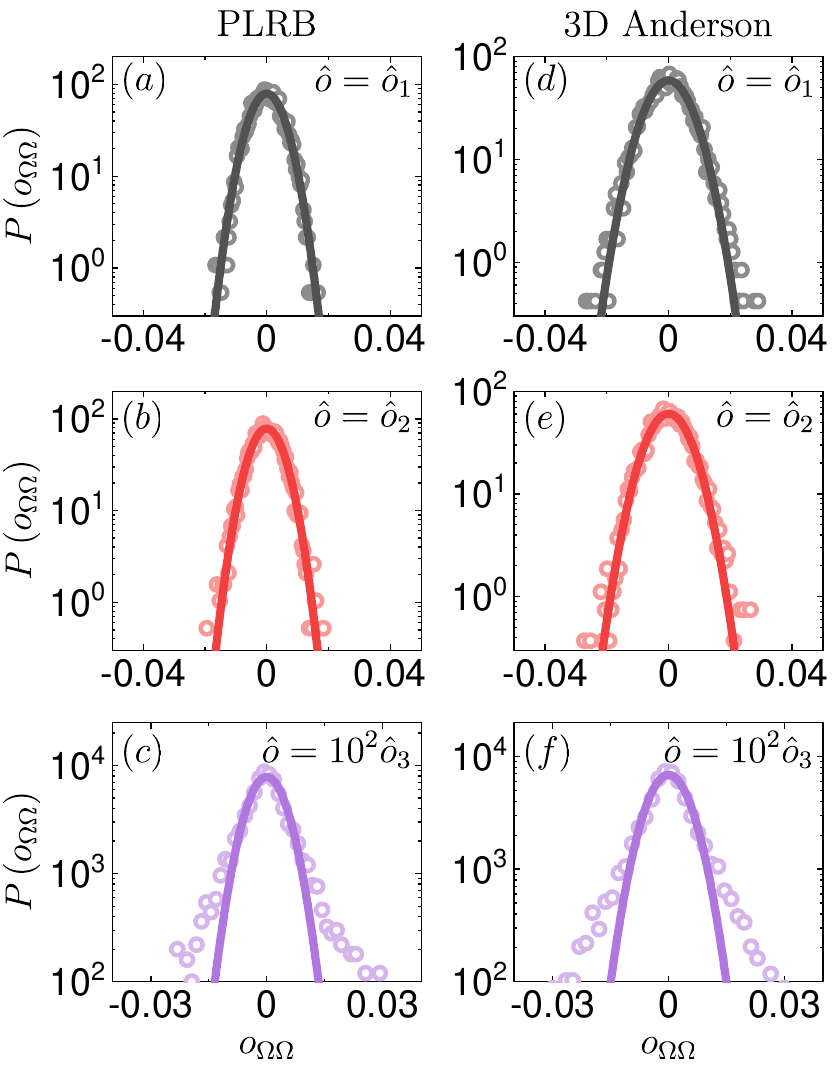}
\vspace{-0.2cm}
\caption{Distributions of diagonal matrix elements of (a),(d) $\hat{o}_{1}$, (b),(e) $\hat{o}_{2}$, and (c),(f) $\hat{o}_{3}$ in (a)--(c) the PRBM model with $\alpha=0.2$ and $\beta=0.1$, and (d)--(f) the 3D Anderson model with $W=5$. We consider $V=10648$ ($V=22^3=10648$) lattice sites and $200$ diagonal matrix elements for $20$ realizations of the PRBM (3D Anderson) model. The lines show least-squares fits to a Gaussian function.}
\label{fig_2body_dist}
\end{figure}

Our results for two-body observables in QCQ models are consistent with the results obtained for the SYK2 model (which is also a QCQ model) in Ref.~\cite{haque_mcclarty_19}. We note that in that study the variance scaled as $\sigma^2 \propto 1/V^2$. This is consistent with Eq.~(\ref{twobody_scenario_2}), and follows from the fact that the diagonal matrix elements of the one-body observables appearing in the Wick's decomposition have a vanishing mean value in the many-body eigenstates.

In Fig.~\ref{fig_2body_dist} we plot the probability density functions (PDFs) of the diagonal matrix elements in the PRBM model (with $\alpha=0.2$ and $\beta=0.1$, left column) and in the 3D Anderson model (with $W=5$, right column), respectively, for the two-body observables considered in Fig.~\ref{fig_anderson_plrb_twobody}. For all two-body observables in the two QCQ models the distributions are close to Gaussian.

\begin{figure}[!t]
\includegraphics[width=0.98\columnwidth]{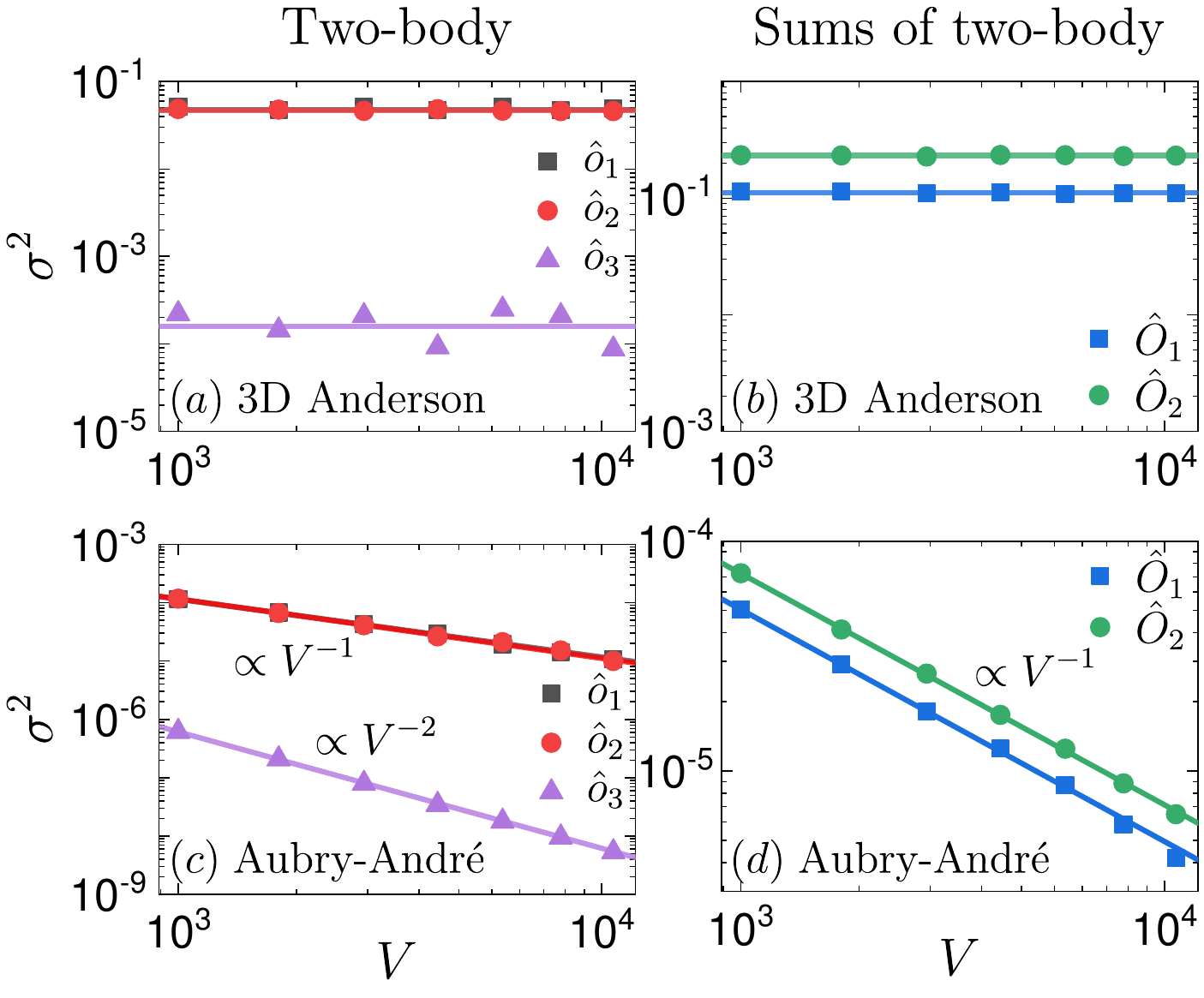}
\vspace{-0.2cm}
\caption{Same as Fig.~\ref{fig_anderson_plrb_twobody} for the (a),(b) 3D Anderson model with $W=30$, and for the (c),(d) 1D Aubry–André model with $\lambda = 0.5$.}
\label{fig_anderson_aa}
\end{figure}

In Fig.~\ref{fig_anderson_aa} we show results for the variances of the diagonal matrix elements of the same observables as in Fig.~\ref{fig_anderson_plrb_twobody} but for localized quadratic Hamiltonians, i.e., for the 3D Anderson model with $W=30$ and for the 1D Aubry–André model with $\lambda=0.5$. The two-body observables $\hat o_1$, $\hat o_2$, and $\hat o_3$, which describe interactions and hoppings that are local in position space, exhibit variances in the localized regime of the 3D Anderson model that do not decay with $V$ [Fig.~\ref{fig_anderson_aa}(a)]. On the other hand, these observables exhibit normal weak eigenstate thermalization in the 1D Aubry–André model with $\lambda=0.5$ [Fig.~\ref{fig_anderson_aa}(c)], which is delocalized (localized) in position (quasimomentum) space. The results in Fig.~\ref{fig_anderson_aa}(c) parallel what happens in the quantum-chaotic regimes in Figs.~\ref{fig_anderson_plrb_twobody}(a) and~\ref{fig_anderson_plrb_twobody}(c). Like the two-body observables, the variances of sums of two-body observables $\hat O_1$ and $\hat O_2$, do not exhibit weak eigenstate thermalization in the localized regime of the 3D Anderson model [Fig.~\ref{fig_anderson_aa}(b)] and exhibit normal weak eigenstate thermalization in the 1D Aubry–André model with $\lambda=0.5$ [Fig.~\ref{fig_anderson_aa}(d)]. 

Our results for the 1D Aubry–André model with $\lambda=0.5$ suggest that, for local two-body observables and their sums, single-particle localization in quasimomentum space results in very similar properties of the diagonal matrix elements as single-particle quantum chaos.

\begin{figure*}[!t]
    \includegraphics[width=0.98\textwidth]{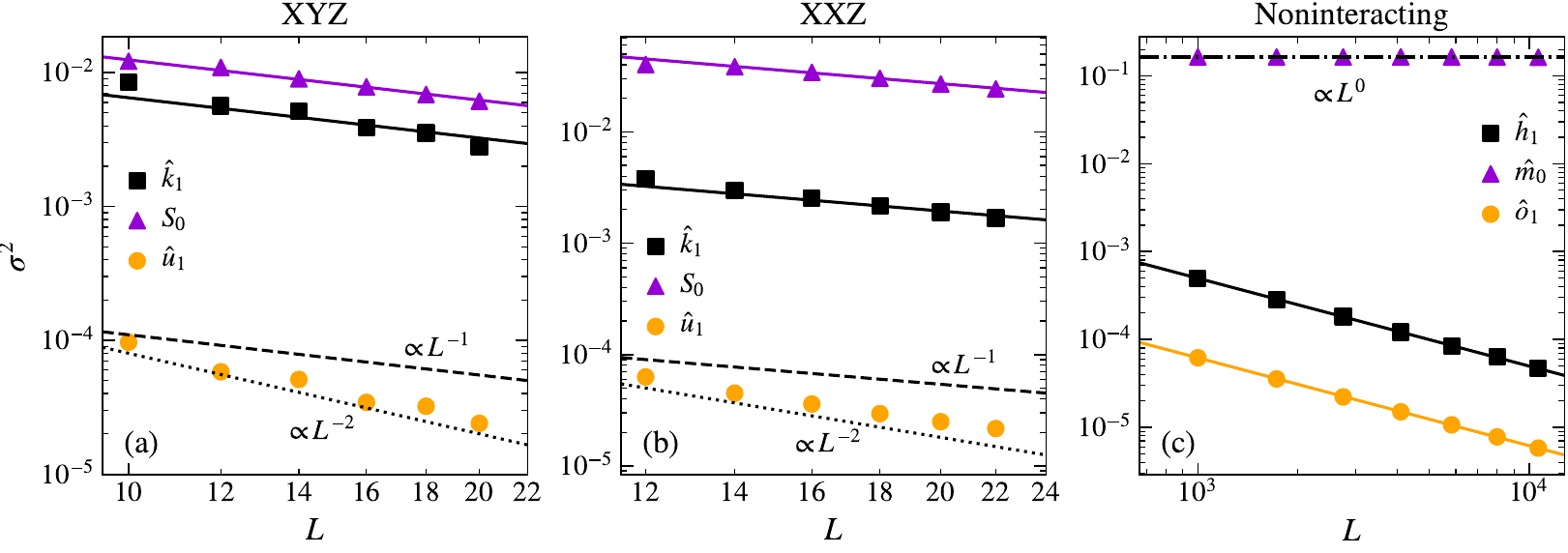}
    \vspace{-0.2cm}
    \caption{Variances of the diagonal matrix elements of few-body observables for the (a) XYZ model, (b) XXZ model, and (c) noninteracting fermions with open boundary conditions. For (a),(b) we calculate the variance within $50\%$ of the energy eigenstates about the mean energy from each symmetry sector, while for the noninteracting fermions in (c) we use the same method as for the other quadratic models. The solid lines show one parameter fits $a/L$ to the last four points of each corresponding data set. The dotted, dashed, and dashed-dotted lines show $L^{-2}$, $L^{-1}$, and $L^0$ functions, respectively, and are guides to the eye.}
    \label{fig_integrable}
\end{figure*}

\section{Normal weak eigenstate thermalization: Integrable interacting models}\label{sec:interacting}

In this section, we study the fluctuations of the diagonal matrix elements of various normalized observables in the eigenstates of paradigmatic integrable interacting spin-$\frac{1}{2}$ chains. We consider the spin-$\frac{1}{2}$ XYZ chain with open boundary conditions with the Hamiltonian
\begin{equation}
\label{eqXYZ}
H_\text{XYZ}=\sum_{\ell=1}^{L-1}\qty(1+\eta)\hat{s}_\ell^x\hat{s}_{\ell+1}^x+\qty(1-\eta)\hat{s}_\ell^y\hat{s}_{\ell+1}^y+\Delta\hat{s}_\ell^z\hat{s}_{\ell+1}^z,
\end{equation}
where $\hat{s}_\ell^\alpha=\frac{1}{2}\sigma^\alpha_\ell$ are the spin-$\frac{1}{2}$ operators at site $\ell$, $\alpha=x,y,z$, and $\sigma^\alpha$ are the Pauli matrices. We also consider the spin-$\frac{1}{2}$ XXZ chain with open boundary conditions with the Hamiltonian
\begin{equation}
H_\text{XXZ}=\sum_{\ell=1}^{L-1}\hat{s}_\ell^x\hat{s}_{\ell+1}^x+\hat{s}_\ell^y\hat{s}_{\ell+1}^y+\Delta\hat{s}_\ell^z\hat{s}_{\ell+1}^z\,.
\end{equation}
The latter model has $U(1)$ symmetry and is obtained from Eq.~(\ref{eqXYZ}) by setting $\eta=0$. We focus on the zero magnetization sector of the spin-$\frac{1}{2}$ XXZ chain. 

We fix the parameters for the XYZ (XXZ) chain to be $\eta=\Delta=0.2$ ($\Delta=0.55$) as in Ref.~\cite{swietek_kliczkowski_24}. We find the energy eigenstates and the corresponding matrix elements of observables carrying out exact diagonalization calculations for chains with even numbers of sites. To this end, we explicitly take into account all the symmetries relevant to each model. Specifically, for both models we account for the reflection and parity, $\mathcal{P}^\alpha=\prod_\ell\sigma^\alpha_\ell$ with $\alpha=x,\,y,\,z$, symmetries. For the XYZ chain with $L$ even, the three $\mathcal{P}^\alpha$ operators commute with each other. Because of the Clifford algebra of the Pauli matrices, one can write any Pauli matrix as the product of the remaining two. We thus need to resolve only two of the parity symmetries (in addition to the reflection symmetry)~\cite{swietek_kliczkowski_24}. For simplicity we choose $\mathcal{P}^x$ and $\mathcal{P}^z$. Therefore, the Hilbert space dimensions for the symmetry blocks considered for the XYZ model scale as $2^L/8$. For the XXZ chain at zero magnetization, $\mathcal{P}^z$ acts trivially (it counts the number of up spins), so we only need to account for $\mathcal{P}^x$ (in addition to the reflection symmetry). Therefore, the Hilbert space dimensions for the symmetry blocks considered scale as $\binom{L}{L/2}/4$.

Our spin-$\frac{1}{2}$ models can be rewritten in terms of hard-core boson operators, and can be mapped onto interacting spinless fermions models by means of the Jordan-Wigner transformation. Using the raising and lowering operators $\hat{s}^{\pm}_\ell=\hat{s}^x_\ell\pm i\hat{s}^y_\ell$, which are equivalent to the creation and annihilation operators for hard-core bosons, and $\hat{s}^{z}_\ell$ operators, which are equivalent to site occupation operators (up to a $1/2$ offset), we define spin-$\frac{1}{2}$ observables that parallel the one- and two-body observables studied for the quadratic spinless fermion models. However, because of the need to carry out full exact diagonalization calculations for the spin chains, the system sizes that can be solved are much smaller than for quadratic models, so our results suffer from stronger finite size effects. Therefore, we focus on the nearest-neighbor one- and two-body operators at the center of the chains. They suffer from weaker finite-size effects than next-nearest neighbor and longer-range operators, and than observables that are sums of few-body operators.

We consider the (one-body) nearest-neighbor ``hopping'' operator
\begin{equation} \label{def_K_l}
    \hat k_1=\hat{s}^+_{L/2}\hat{s}^-_{L/2+1}+\hat{s}^+_{L/2+1}\hat{s}^-_{L/2}\,,
\end{equation}
the (two-body) nearest-neighbor ``density-density'' operator 
\begin{equation} \label{def_U_l}
    \hat u_1=\hat{s}^z_{L/2}\hat{s}^z_{L/2+1} \,,
\end{equation}
and the equivalent of the (one-body) occupation of the zero quasimomentum mode
\begin{equation}\label{def_S0}
    \hat S_0=\frac{1}{L}\sum_{\ell,\ell'=1}^L\hat{s}^+_\ell\hat{s}^-_{\ell'}\,.
\end{equation}

The scaling of variances of the diagonal matrix elements of the observables in Eqs.~(\ref{def_K_l})-(\ref{def_S0}) are shown in Fig.~\ref{fig_integrable} for the XYZ [Fig.~\ref{fig_integrable}(a)] and the XXZ [Fig.~\ref{fig_integrable}(b)] models, together with the results for $\hat h_1$, $\hat o_1$, and $\hat m_0$ for spinless fermions with nearest neighbor hoppings in a chain with open boundary conditions [Fig.~\ref{fig_integrable}(c)]. Like for the Aubry–André model with $\lambda = 0.5$ in Figs.~\ref{fig_aa}(a) and~\ref{fig_anderson_aa}(c), in Fig.~\ref{fig_integrable}(c) one can see that $\sigma^2(\hat h_1)\propto 1/L$, $\sigma^2(\hat o_1)\propto 1/L$, and $\sigma^2(\hat m_0)\propto L^0$. In the spin chains [Figs.~\ref{fig_integrable}(a) and~\ref{fig_integrable}(b)], $\sigma^2(\hat k_1)\propto 1/L$, $\sigma^2(\hat u_1)\propto 1/L^\alpha$ with $\alpha\in[1,2]$, and $\sigma^2(\hat S_0)\propto 1/L$. Therefore, because of the presence of interactions, in the clean spin chains $\hat S_0$ exhibits normal weak eigenstate thermalization. This is to be contrasted to the fact that the corresponding fermionic operator $\hat m_0$ does not exhibit normal weak eigenstate thermalization for spinless fermions with nearest neighbor hoppings in a chain with open boundary conditions. The latter is a consequence of single-particle localization in quasimomentum space. 

Consistent with our results that normal weak eigenstate thermalization occurs for $\hat S_0$ in the XYZ and XXZ chains, in Ref.~\cite{zhang_vidmar_22} it was shown that the variance of $\hat m_0$ is $\propto 1/V$ in a translationally invariant model of hard-core bosons with nearest neighbor hoppings. Again, in such a clean model, interactions are responsible for the occurrence of normal weak eigenstate thermalization for $\hat m_0$, like single-particle eigenstate thermalization ensures that normal weak eigenstate thermalization occurs for one-body observables in QCQ models. 

\section{Summary and discussion} \label{sec:conclusions}

\begin{figure}[!t]
\includegraphics[width=1\columnwidth]{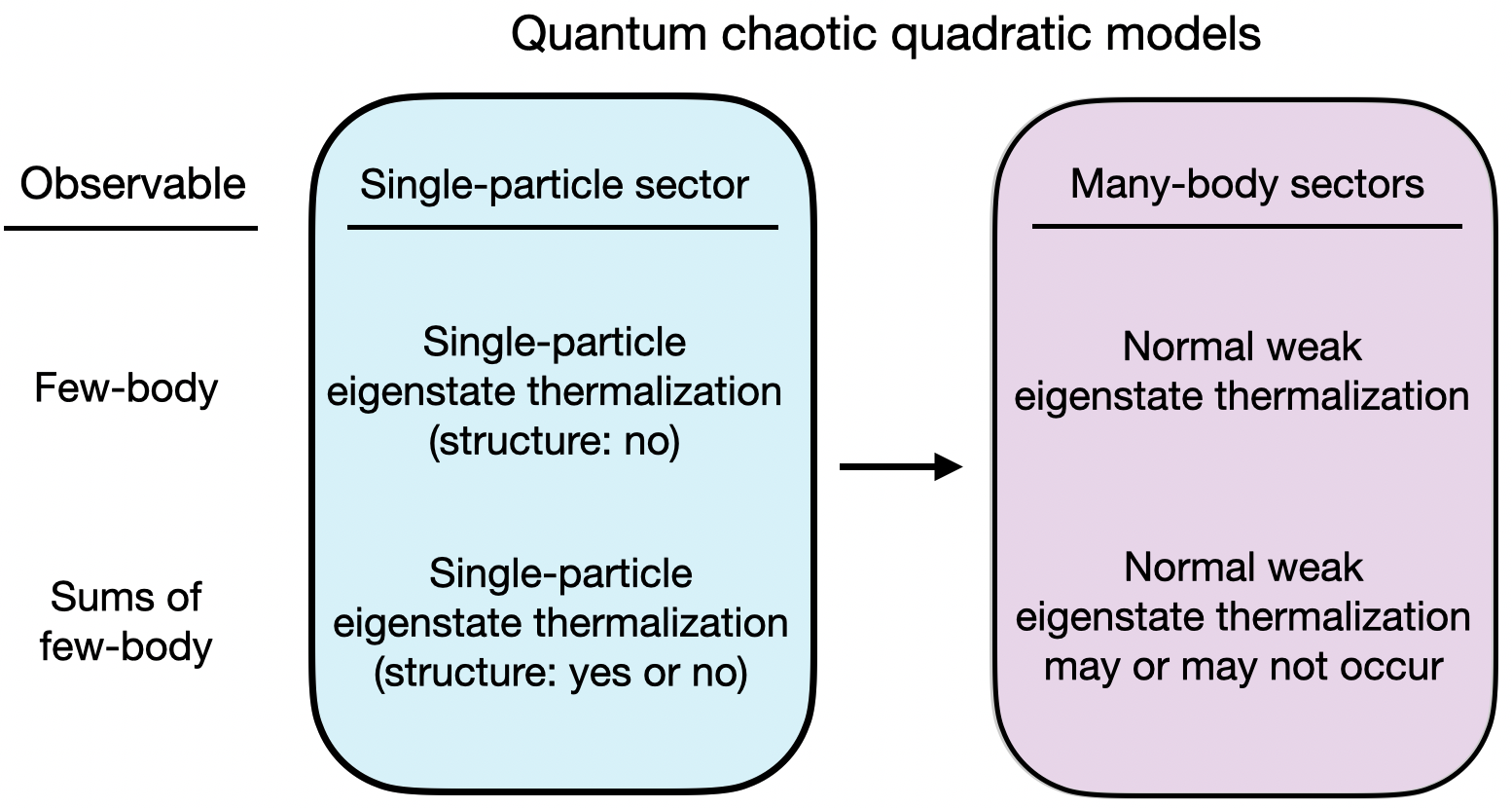}
\caption{Summary of our results for the fluctuations of the diagonal matrix elements of observables in the single-particle and many-body sectors of quantum-chaotic quadratic (QCQ) Hamiltonians. The upper row applies to the few-body (one- and two-body) observables introduced in Sec.~\ref{sec:fewbody}, while lower row applies to the sums of few-body observables introduced in Sec.~\ref{sec:sums}. Our main result is that eigenstate thermalization in the single-particle sector ensures normal weak eigenstate thermalization of one- and two-body observables in the many-body eigenstates. On the other hand, for sums of one-body observables the structure of the observable in the single-particle energy eigenstates may preclude normal weak eigenstate thermalization in the many-body eigenstates. The fluctuations of the diagonal matrix elements of few-body observables in many-body eigenstates of integrable interacting models were found to be qualitatively similar to those in QCQ models.}
\label{fig0}
\end{figure}

We proved that single-particle eigenstate thermalization guaranties the occurrence of normal weak eigenstate thermalization of one-body observables, and provided numerical evidence of this phenomenon in QCQ models like the 3D Anderson and the PLRB models in the delocalized regime. We provided analytical arguments and numerical results that suggest that normal weak eigenstate thermalization also occurs in those cases for two-body observables, so we expect our conclusions to apply to few-body observables in general. By normal weak eigenstate thermalization we mean that the variance of the diagonal matrix elements of normalized observables calculated over the entire spectrum of many-body energy eigenstates decays polynomially or faster with increasing system size. On the other hand, sums of one- and two-body observables may or may not exhibit normal weak eigenstate thermalization. For sums of one-body observables, we traced the possible behaviors of the variance to the structure of the operators in the single-particle sector. We presented numerical evidence supporting those results in various QCQ models, and our results are summarized in Fig.~\ref{fig0}. We also showed that normal weak eigenstate thermalization fails to occur for one- and two-body observables that are local in the space in which localization occurs in quadratic models that are not quantum chaotic.

For integrable interacting models, we provided numerical evidence that normal weak eigenstate thermalization occurs for all the one- and two-body observables considered. Our findings suggest that in those models interactions effectively produce the same effect in the one- and two-body sectors of the many-body eigenstates (after tracing out all but one or two particles, respectively) that single-particle quantum chaos produces in QCQ models. This is remarkable in the light of the fundamental differences between the off-diagonal matrix elements (the ones responsible for the quantum dynamics~\cite{dalessio_kafri_16}) of few-body observables in integrable interacting and quadratic models. The off-diagonal matrix elements in many-body eigenstates of quadratic models are sparse~\cite{khatami_pupillo_13, haque_mcclarty_19, zhang_vidmar_22} and the magnitude of the nonzero ones decays polynomially with the system size~\cite{zhang_vidmar_22}. In integrable interacting models, on the other hand, the off-diagonal matrix elements are dense and their magnitude decays exponentially with increasing system size~\cite{leblond_mallayya_19, zhang_vidmar_22}. 

Our results suggests that QCQ models can be used to gain analytical insights into the behavior of the diagonal matrix elements of few-body observables in interacting integrable models and, potentially, on that of few-body observables that are nonlocal in the space in which localized quadratic models exhibit single-particle localization. Some important open questions still remain in the context of QCQ models. For example, we still need to understand the conditions needed for normal weak eigenstate thermalization to occur for sums of two and higher few-body observables, and explore what happens when the variances are calculated in microcanonical energy windows away from the center of the spectrum. For quantum dynamics, interesting open questions include: (i) are there classes of physically relevant initial states for which observables that exhibit normal weak eigenstate thermalization thermalize under noninteracting or integrable interacting dynamics and, (ii) for more general initial states, can one create nonthermal states after equilibration that are useful for quantum information storage and/or computing.

\acknowledgements
We acknowledge discussions with M. Hopjan. This work was supported by the Slovenian Research and Innovation Agency (ARIS), Research core funding Grants No.~P1-0044, N1-0273, J1-50005 and N1-0369 (R.Ś. and L.V.), and by the National Science Foundation under Grant No.~PHY-2309146 (M.R.). Part of the numerical studies in this work were carried out using resources provided by the Wroclaw Centre for Networking and Supercomputing~\cite{wcss}, Grant No.~579 (P.Ł.). We gratefully acknowledge the High Performance Computing Research Infrastructure Eastern Region (HCP RIVR) consortium~\cite{vega1} and European High Performance Computing Joint Undertaking (EuroHPC JU)~\cite{vega2}  for funding this research by providing computing resources of the HPC system Vega at the Institute of Information sciences~\cite{vega3}.

\appendix

\section{Absence of structure for one-body observables in single-particle eigenstates} \label{sec:app_no_structure}

We prove that one-body observables in the single-particle energy eigenstates do not have structure in the thermodynamic limit, i.e., that $\oo(\epsilon_\omega)$ in Eq.~\eqref{O_sp} is a constant. Specifically, we prove that when writing
\begin{equation} \label{def_structure_polynomial:app}
    \oo(\epsilon_\omega)=\sum_{i=0}^{V} p_{i}(\epsilon_\omega)^i\,,
\end{equation}
where $\epsilon_\omega$ are the single-particle energies, the coefficients $p_i$ vanish for all $i \geq 1$.

Normalized one-body observables $\hat{o}/||\hat{o}||_\text{sp}$ can be written as~\cite{lydzba_zhang_21}
\begin{equation} \label{def_onebody_general}
    \frac{\hat{o}}{||\hat{o}||_\text{sp}}=\mathbf{O}\left(\sqrt{V}\right)\sum_{i=1}^{\mathbf{O}(1)}o_i\hat{f}^\dagger_i\hat{f}^{}_{i}\,,
\end{equation}
where $\hat{f}^\dagger_i$ ($\hat{f}^{}_i$) creates (annihilates) a spinless fermion in the $i^{\rm th}$ single-particle eigenstate of $\hat{o}$, and $\sum_{i=1}^{\mathbf{O}(1)}o_i$ is at most $\mathbf{O}(1)$. The diagonal matrix elements $o_{\omega\omega} = \langle\omega|\hat{o}|\omega\rangle$ in the single-particle energy eigenstates $|\omega\rangle$ are
\begin{equation}
    \frac{o_{\omega\omega}}{||\hat{o}||_\text{sp}} = 
    \mathbf{O}\left(\sqrt{V}\right)\sum_{i=1}^{\mathbf{O}(1)} o_i |\langle \omega|i \rangle|^2\,,
\end{equation}
where $\langle\omega|\hat f_i^\dagger \hat f^{}_i |\omega\rangle = |\langle\omega|i\rangle|^2$.

Next, we assume that the structure in the single-particle sector can be determined after coarse-graining all single-particle energy eigenstates into an arbitrary large but ${\bf O}(1)$ number of bins, each containing ${\bf O}(V)$ eigenstates, and averaging the diagonal matrix elements within each bin. Specifically, we define an energy window $\Delta$ that comprises ${\cal N} = {\bf O}(V)$ single-particle energy eigenstates from an energy bin with the mean energy $\overline{\epsilon} = \sum_{\omega\in\Delta}\epsilon_\omega/{\cal N}$. The average value of the observable at this energy is then
\begin{equation}
    \oo(\overline{\epsilon})= \frac{1}{\mathbf{O}(V)}\sum_{\omega\in\Delta}\mathbf{O}\left(\sqrt{V}\right)\sum_{i=1}^{\mathbf{O}(1)} o_i |\langle \omega|i \rangle|^2 \,.
\end{equation}
Since $\sum_{\omega=1}^{V}|\langle\omega|i\rangle|^2=1$, we have that $\sum_{\omega\in\Delta}|\langle\omega|i\rangle|^2$ is at most $\mathbf{O}(1)$, and the structure function for traceless observables is upper bounded as
\begin{equation}
    \oo(\overline{\epsilon}) \leq \mathbf{O}(1/\sqrt{V})\,,
\end{equation}
i.e., the coefficients $p_i$ in the polynomial expansion in Eq.~(\ref{def_structure_polynomial:app}) are at most ${\bf O}(1/\sqrt{V})$ and so they vanish in the thermodynamic limit. 

For extensive sums of one-body observables, the equivalent of Eq.~(\ref{def_onebody_general}) is~\cite{lydzba_zhang_21}
\begin{equation}
     \frac{\hat{o}}{||\hat{o}||_\text{sp}}=\sum_{i=1}^{\mathbf{O}(V)} o_i\hat{f}^\dagger_i\hat{f}_{i} \,.
\end{equation}
Repeating the same steps as for one-body observables, one obtains that
\begin{equation}
    \oo(\overline{\epsilon}) \leq \mathbf{O}(1)\,,
\end{equation}
i.e., the structure of sums of one-body observables in the single-particle sector may not vanish. Hence, there may exist $p_i$ (with $i \geq 1$) in the polynomial expansion in Eq.~(\ref{def_structure_polynomial:app}) that are ${\bf O}(1)$.

\section{Variance of one-body observables} \label{app:variance}

We calculate the variances of $\chi_{1,\Omega}$ and $\chi_{2,\Omega}$ in Eqs.~(\ref{def_chi_1}) and~(\ref{def_chi_2}), respectively. For simplicity, the variances ${\rm Var}(...)$ and the means $\mathbb{E}[...]$ are calculated over the entire many-body Hilbert space $D$ including all particle number sectors $N$. Since the numerical computations are carried out at a fixed particle filling $n=N/V=1/2$, at the end of each analytic calculation we mention how the results are modified when $n$ is fixed.

\subsection{Variance of $\chi_{2,\Omega}$} \label{app:variance_chi_2}

First, we calculate the variance of $\chi_{2,\Omega}$ discussed in Sec.~\ref{sec:argument}. We rewrite $\chi_{2,\Omega}$ in Eq.~(\ref{def_chi_2}) as
\begin{equation} \label{eq:chi2_def2}
    \chi_{2,\Omega} = ||\hat o||_{\rm sp} \sum_{\omega=1}^V A_\omega \, \langle \Omega| \hat f_\omega^\dagger \hat f_\omega |\Omega\rangle \,,
\end{equation}
so that the mean of $\chi_{2,\Omega}$ can be expressed as
\begin{equation} \label{eq:chi2_avr}
    \mathbb{E}[\chi_{2,\Omega}] = V ||\hat o||_{\rm sp} n \, \bar A_{\rm sp}\,,
\end{equation}
where $n$ is the average occupation of single-particle energy eigenstates in the many-body energy eigenstates, i.e., the average filling in the many-body Hilbert space. In Eqs.~(\ref{eq:chi2_def2}) and~(\ref{eq:chi2_avr}), we defined the quantity $A_\omega$ and its average $\bar A_{\rm sp}$ in the single-particle Hilbert space as
\begin{equation}
    A_\omega= \frac{{\cal F}_o (\epsilon_\omega,0)}{\sqrt{\rho(\epsilon_\omega)}} R_{\omega\omega}^o \,,\,\,\,
    \bar A_{\rm sp} = \frac{1}{V} \sum_{\omega=1}^V \frac{{\cal F}_o (\epsilon_\omega,0)}{\sqrt{\rho(\epsilon_\omega)}} R_{\omega\omega}^o \,.
\end{equation}
Since $\bar A_{\rm sp} \to 0$ for sufficiently large $V$, we have that $\mathbb{E}[\chi_{2,\Omega}] \to 0$. The variance can then be calculated as
\begin{align}
    &{\rm Var}[\chi_{2,\Omega}] = ||\hat o||_{\rm sp}^2 \frac{1}{D} \sum_{\Omega=1}^D \left( \sum_{\omega=1}^V A_\omega \, \langle \Omega| \hat f_\omega^\dagger \hat f^{}_\omega |\Omega\rangle \right)^2 \nonumber \\
    & = ||\hat o||_{\rm sp}^2 \sum_{\omega=1}^V (A_\omega)^2 \frac{1}{D} \sum_{\Omega=1}^D \langle\Omega| \hat f_\omega^\dagger \hat f^{}_\omega|\Omega\rangle^2  \label{eq:chi2_var1} \\
    & + ||\hat o||_{\rm sp}^2 \sum_{\omega\neq\omega'=1}^V A_\omega A_{\omega'} \frac{1}{D} \sum_{\Omega=1}^D \langle\Omega| \hat f_\omega^\dagger \hat f^{}_\omega|\Omega\rangle \langle\Omega|\hat f_{\omega'}^\dagger \hat f^{}_{\omega'}|\Omega\rangle\,. \nonumber
\end{align}
The averages over the many-body Hilbert space dimension $D$ in Eq.~(\ref{eq:chi2_var1}) equal $n$ and $n^2$, respectively, and hence we obtain
\begin{align}
    & {\rm Var}[\chi_{2,\Omega}] = ||\hat o||_{\rm sp}^2 n \left(\sum_{\omega=1}^V (A_\omega)^2 \right) \nonumber \\
    & \hspace*{2.0cm} + ||\hat o||_{\rm sp}^2 n^2 \left[ \left(\sum_{\omega=1}^V A_\omega \right)^2 - \sum_{\omega=1}^V (A_\omega)^2 \right] \nonumber \\
    & \hspace*{1.5cm} = V ||\hat o||_{\rm sp}^2 n(1-n) \overline {A^2}_{\rm sp} \,, \label{eq:chi2_var2}
\end{align}
which is the result reported in Eq.~(\ref{def_variance_chi2}). This result is independent of whether the many-body Hilbert space, of dimension $D$, includes all particle-number sectors or it is limited to a fixed particle-number sector.

\subsection{Variance of $\chi_{1,\Omega}$} \label{app:variance_chi_1}

Next, we calculate the variance of $\chi_{1,\Omega}^{(i)}$ in Eq.~(\ref{def_chi_1_omega_i}) for $i\geq 2$. (Recall that, for $i=0$ and $1$, the contributions to $\chi_{1,\Omega}$ are smooth functions of the energy.) For $i\geq 2$, the mean of $\chi_{1,\Omega}^{(i)}$ reads
\begin{equation}
    \mathbb{E}[\chi_{1,\Omega}^{(i)}] = V ||\hat o||_{\rm sp} n\, p_i\, \overline{\epsilon^i}_{\rm sp} \,,
\end{equation}
where the mean of the single-particle energies to the $i$-th power is defined as
\begin{equation} \label{def_epsilon_avr}
    \overline{\epsilon^i}_{\rm sp} = \frac{1}{V} \sum_{\omega=1}^V (\epsilon_\omega)^i \,.
\end{equation}
We can write the variance of $\chi_{1,\Omega}^{(i)}$ as
\begin{align}
    {\rm Var}[\chi_{1,\Omega}^{(i)}] &= ||\hat o||_{\rm sp}^2\, p_i^2\,
    \frac{1}{D} \sum_{\Omega=1}^D \left( \sum_{\omega=1}^V (\epsilon_\omega)^i \, \langle \Omega| \hat f_\omega^\dagger \hat f^{}_\omega |\Omega\rangle \right)^2 \nonumber \\
    & - \mathbb{E}^2[\chi_{1,\Omega}^{(i)}] \,.
\end{align}
The steps to simplify this expression are analogous to those in Eqs.~(\ref{eq:chi2_var1})-(\ref{eq:chi2_var2}) to obtain the variance of $\chi_{2,\Omega}$, and they yield
\begin{equation} \label{eq:chi1_var2}
    {\rm Var}[\chi_{1,\Omega}^{(i)}] = V ||\hat o||_{\rm sp}^2 n(1-n) \, p_i^2 \, \overline{\epsilon^{2i}}_{\rm sp} \,,
\end{equation}
which is the result reported in Eq.~(\ref{def_variance_chi1}).

Equation~\eqref{def_variance_chi1} was obtained by carrying out the average over the many-body Hilbert space $D$ including all particle-number sectors. If one restricts the calculation to a sector with a fixed filling $n$ (we set $n=1/2$ in the numerical calculations), Eq.~(\ref{eq:chi1_var2}) is modified to read
\begin{equation} \label{eq:chi1_var2_v2}
    {\rm Var}[\chi_{1,\Omega}^{(i)}] =  V ||\hat o||_{\rm sp}^2 n(1-n) \, p_i^2 \, \left[ \overline{\epsilon^{2i}}_{\rm sp} - \left(\overline{\epsilon^{i}}_{\rm sp}\right)^2 \right] \,.
\end{equation}
The modification does not change the order of scaling of the variance with $V$, i.e., it does not affect whether normal weak eigenstate thermalization occurs or not. Note that the difference between Eqs.~(\ref{eq:chi1_var2}) and~(\ref{eq:chi1_var2_v2}) arises from the fact that the particle-number variance at infinite temperature is different in the grand canonical ensemble compared to the canonical ensemble.

\section{Variance of products of site occupations} \label{app:2body_arguments}

Building on the analysis in Sec.~\ref{sec:beyond_onebody}, we detail our argument for the scaling of variances of two-body observables for the particular case in which they are products of site occupations, $\hat n_i \hat n_j$, with $i \neq j$. $\hat{o}_{1}$ and $\hat{o}_{2}$ in Eqs.~(\ref{eq_o1}) and~(\ref{eq_o2}), respectively, are examples of such observables.

The diagonal matrix elements of $\hat n_i \hat n_j$ can be expressed using the Wick decomposition [see Eq.~(\ref{def_wick})] as
\begin{equation}
    (n_{i} n_{j})_{\Omega\Omega} = 
    (n_{i})_{\Omega\Omega} (n_{j})_{\Omega\Omega} -
    |(c_{i}^\dagger c^{}_{j})_{\Omega\Omega}|^2,
\end{equation}
where $(n_{i} n_{j})_{\Omega\Omega}=\langle\Omega|\hat n_{i} \hat n_{j}|\Omega\rangle$, etc, and we assumed that $|\Omega\rangle$ is a Gaussian state with a fixed particle number (an eigenstate of $\hat N=\sum_{i=1}^{V}\hat{n}_{i}$). The variance of $(n_{i} n_{j})_{\Omega\Omega}$ then satisfies the inequality from Eq.~(\ref{covariance_inequality}),
\begin{equation}
\label{eq_app_sigma}
    \sigma^2\left[ (n_{i} n_{j})_{\Omega\Omega} \right] \le \left( \sigma[(n_{i})_{\Omega\Omega} (n_{j})_{\Omega\Omega}] + \sigma[ |(c_{i}^\dagger c^{}_{j})_{\Omega\Omega}|^2 ]\right)^2\,.
\end{equation}

The square of the first term in the parenthesis in Eq.~(\ref{eq_app_sigma}) can be expressed as
\begin{equation}
\begin{split}
    \sigma^2 [(n_{i})_{\Omega\Omega} (n_{j})_{\Omega\Omega}] 
    & = \sigma^2\left[ (n_{i})_{\Omega\Omega} \right]
    \sigma^2\left[ (n_{j})_{\Omega\Omega} \right] \\ 
    & \hspace*{-2.0cm} + \sigma^2\left[ (n_{i})_{\Omega\Omega} \right] \mathbb{E}^2[ (n_{j})_{\Omega\Omega} ]
    + \sigma^2\left[ (n_{j})_{\Omega\Omega} \right] \mathbb{E}^2[ (n_{i})_{\Omega\Omega} ].
\end{split}
\end{equation}
If both one-body observables $\hat{n}_{i}$ and $\hat{n}_{j}$ exhibit single-particle eigenstate thermalization, we can estimate the scaling of this variance as
\begin{equation} \label{eq:sigma2nn_first}
\begin{split}
     \sigma^2 [(n_{i})_{\Omega\Omega} (n_{j})_{\Omega\Omega}]  & =\textbf{O}(1/V)\textbf{O}(1/V) + \textbf{O}(1/V) \mathbb{E}^2[ (n_{j})_{\Omega\Omega} ] \\
     & +
     \textbf{O}(1/V) \mathbb{E}^2[ (n_{i})_{\Omega\Omega} ] = \textbf{O}(1/V)\,,
\end{split}
\end{equation}
where we have used the fact that $\mathbb{E}^2[ (n_{j})_{\Omega\Omega} ]$ and $\mathbb{E}^2[ (n_{i})_{\Omega\Omega} ]$ are $\textbf{O}(1)$.

In the second term in the parenthesis in Eq.~(\ref{eq_app_sigma}), assuming that the matrix element of the non-Hermitian operator $(c_{i}^\dagger c^{}_{j})_{\Omega\Omega}$ is real, which is the case for all quadratic Hamiltonians considered here, we replace it with a Hermitian operator, $(c_{i}^\dagger c^{}_{j} + c_{j}^\dagger c^{}_{i})_{\Omega\Omega}/2$.
Therefore, the corresponding variance can be expressed as
\begin{equation}
 \begin{split}
 \label{eq_app_term2}
    \sigma^2[ |(c_{i}^\dagger c^{}_{j})_{\Omega\Omega}|^2 ] & \propto 
    \sigma^4[ (c_{i}^\dagger c^{}_{j} + c_{j}^\dagger c^{}_{i})_{\Omega\Omega} ] \\
    & + 2\sigma^2[ (c_{i}^\dagger c^{}_{j} + c_{j}^\dagger c^{}_{i})_{\Omega\Omega} ] \,
    \mathbb{E}^2[(c_{i}^\dagger c^{}_{j} + c_{j}^\dagger c^{}_{i})_{\Omega\Omega} ].
 \end{split}
\end{equation}
Assuming again that the one-body observable $\hat{c}_{i}^\dagger\hat{c}^{}_{j}+\hat{c}_{j}^\dagger\hat{c}^{}_{i}$ exhibits the single-particle eigenstate thermalization, and exploiting that $\mathbb{E}[(c_{i}^\dagger c^{}_{j} + c_{j}^\dagger c^{}_{i})_{\Omega\Omega} ]=0$, we arrive at
\begin{equation}
 \label{eq_app_term22}
    \sigma^2[ |(c_{i}^\dagger c^{}_{j})_{\Omega\Omega}|^2 ]=\textbf{O}(1/V^2) \,.
\end{equation}

The above analysis shows that the upper bound of the variance of $(n_{i} n_{j})_{\Omega\Omega}$ is set by Eq.~(\ref{eq:sigma2nn_first}), and hence
\begin{equation}
    \sigma^2[ (n_{i} n_{j})_{\Omega\Omega} ] \le \textbf{O}(1/V) \,.
\end{equation}
On the other hand, if for a given two-body observable all one-body operators appearing in the Wick decomposition have a vanishing mean of the diagonal matrix elements (the case for $\hat{c}_{i}^\dagger\hat{c}_{j}+\hat{c}_{j}^\dagger\hat{c}_{i}$), then the upper bound of the variance is $\textbf{O}(1/V^2)$.

\section{Additional results for matrix elements} \label{app:distributions}

\begin{figure}[!h]
\includegraphics[width=0.98\columnwidth]{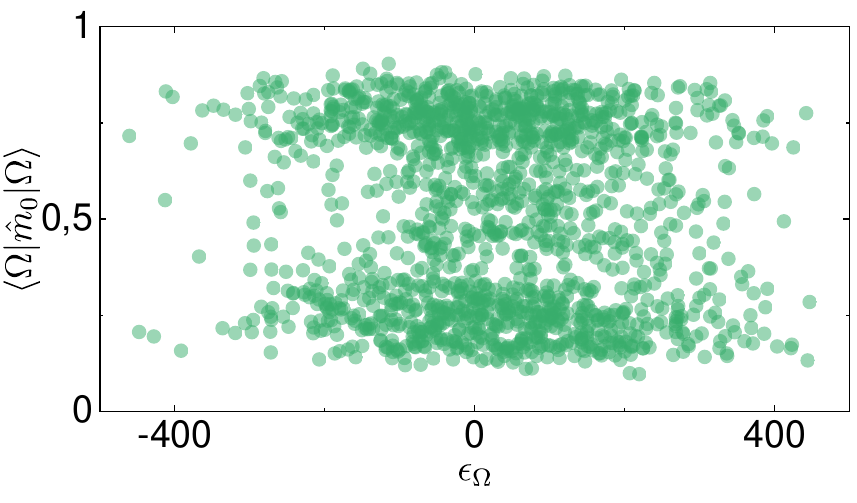}
\vspace{-0.2cm}
\caption{Diagonal matrix elements of $\hat{m}_{0}$ in the 3D Anderson model for $W=5$. The 3D lattice has $V=22^3=10648$ sites, and we report $200$ diagonal matrix elements for $10$ disorder realizations.}
\label{fig_anderson_m0}
\end{figure}

The non-Gaussian PDF of the diagonal matrix elements of $\hat m_0$ for the 3D Anderson model in the single-particle quantum-chaotic regime ($W=5$) can be better understood directly looking at the behavior of the matrix elements. In Fig.~\ref{fig_anderson_m0}, we plot randomly selected matrix elements for a system with $V=22^3$ sites. The matrix elements exhibit clustering at values that are different from the mean, which resembles the results in the translationally-invariant point ($W=0$). At the latter point, the matrix elements only take two values $\pm 1$, depending on whether the single-particle ground state is occupied in a many-body energy eigenstate or not. We expect that, for sufficiently large systems, the matrix elements of $\hat m_0$ for the 3D Anderson model with $W=5$ will be normally distributed about the mean value.

\bibliographystyle{biblev1}
\bibliography{references}

\end{document}